\documentclass[11pt,a4paper]{article}
\pdfoutput=1
\usepackage{jheppub}

\usepackage{hyperref}

\usepackage{physics}
\usepackage{amsfonts}
\usepackage{graphicx}
\usepackage{amsmath}
\usepackage{url}
\usepackage{subcaption}
\def\eps{\ensuremath{\varepsilon}}
\def\HR{\ensuremath{\mathcal{R}}}
\def\HI{\ensuremath{\mathcal{I}}}

\title{Three-loop massive tadpoles and polylogarithms through weight six}

\author{B.~A.~Kniehl,}
\author[1]{A.~F.~Pikelner,\note{On leave of absence from Joint Institute for Nuclear
    Research, 141980 Dubna, Russia.}}
\author{and O.~L.~Veretin}

\affiliation{II Institut f\"ur Theoretische Physik,  Universit\"at Hamburg, \\ Luruper Chaussee 149, 22761 Hamburg, Germany}

\emailAdd{kniehl@desy.de}
\emailAdd{andrey.pikelner@desy.de}
\emailAdd{oleg.veretin@desy.de}

\abstract{
  We evaluate the three-loop massive vacuum bubble diagrams
  in terms of polylogarithms up to weight six.
  We also construct the basis of irrational constants being
  harmonic polylgarithms of arguments $e^{ki\pi/3}$.
}


\preprint{DESY 17-070}

\arxivnumber{1705.05136}

\begin{document}
\maketitle
\flushbottom
\allowdisplaybreaks
\section{Introduction}

More than two decades ago, the integration-by-parts relations \cite{Chetyrkin:1981qh} and
asymptotic expansions \cite{Smirnov:1990rz,Smirnov:1994tg}
became common in the Feynman diagram calculus.
The combination of these methods provides a powerful tool for the evaluation of multiloop diagrams. In particular, massive propagator diagrams through the three-loop
order can be reduced with the help of asymptotic-expansion methods to three-loop massive 
tadpoles, which can be done, e.g., using the \texttt{FORM}~\cite{Vermaseren:1992vn} package \texttt{MATAD}~\cite{Steinhauser:2000ry}
(see also~ref.~\cite{Avdeev:1995eu}).%
\footnote{The up-to-date package
  \texttt{MATAD-ng} with full dependence on $d$ can be downloaded from the URL
  \mbox{\url{https://github.com/apik/matad-ng}} and the results of this paper from the
  direct link \mbox{\url{https://git.io/mtdw6}}.}

There are a lot of physical applications, where the above-mentioned technique was
applied. 
Just to mention but a few examples, it was applied to the evaluation of the three-loop $\rho$ parameter 
in QCD \cite{Avdeev:1994db,Chetyrkin:1995ix} and the electroweak theory \cite{Faisst:2003px},
the three-loop QCD corrections to heavy-quark production \cite{Chetyrkin:1996cf},
and many other quantities. Integral topologies with all lines massive find
applications in calculations of renormalization group
functions \cite{Chetyrkin:1997fm,Bednyakov:2013eba,Chetyrkin:2013wya}
at the three-loop order and also at higher orders of the epsilon expansion in
four-loop \cite{Czakon:2004bu} and even five-loop \cite{Luthe:2017ttc} calculations.

In his work \cite{Broadhurst:1998rz}, Broadhurst noticed that all three-loop single-scale
vacuum diagrams at order $O((4-d)/2)$ in dimensional regularization 
can be related to the elements of the algebra of the sixth 
root of unity. This observation allowed him to evaluate all the three-loop integrals up to their
finite parts in terms of a few constants, being polylogarithms of weight four.

In this paper, we proceed by studying three-loop vacuum integrals with a single mass scale
at weights five and six. On the one hand, this is a necessary ingredient in evaluations
beyond the three-loop approximation, where the three-loop master integrals have to be
expanded to higher powers in $d-4$. On the other hand, we would like to test the basis
of the algebra of the sixth root of unity through weight six.

\section{Notation}

We use dimensional regularization with the dimension of space-time being
$d=4-2\varepsilon$ in euclidean space. Each loop integration is normalized as follows:
\begin{align}
  \int d[k] \dots = e^{\gamma\varepsilon}\int \frac{d^d k}{\pi^{d/2}} \dots \,, 
\end{align}
where $\gamma=0.577216\dots$ is the Euler--Mascheroni constant.

Defining
\begin{equation}
  \label{eq:den-def}  C_{i;m}=k_i^2+m^2, \quad   C_{ij;m}=(k_i-k_j)^2+m^2\,,
\end{equation}
the general three-loop vacuum bubble diagram with six scalar
propagators, where it is implied that the masses either take the values
$m$ or zero, may be written as
\begin{align}
  \label{eq:int-def}
  \vcenter{\hbox{\includegraphics[width=2cm]{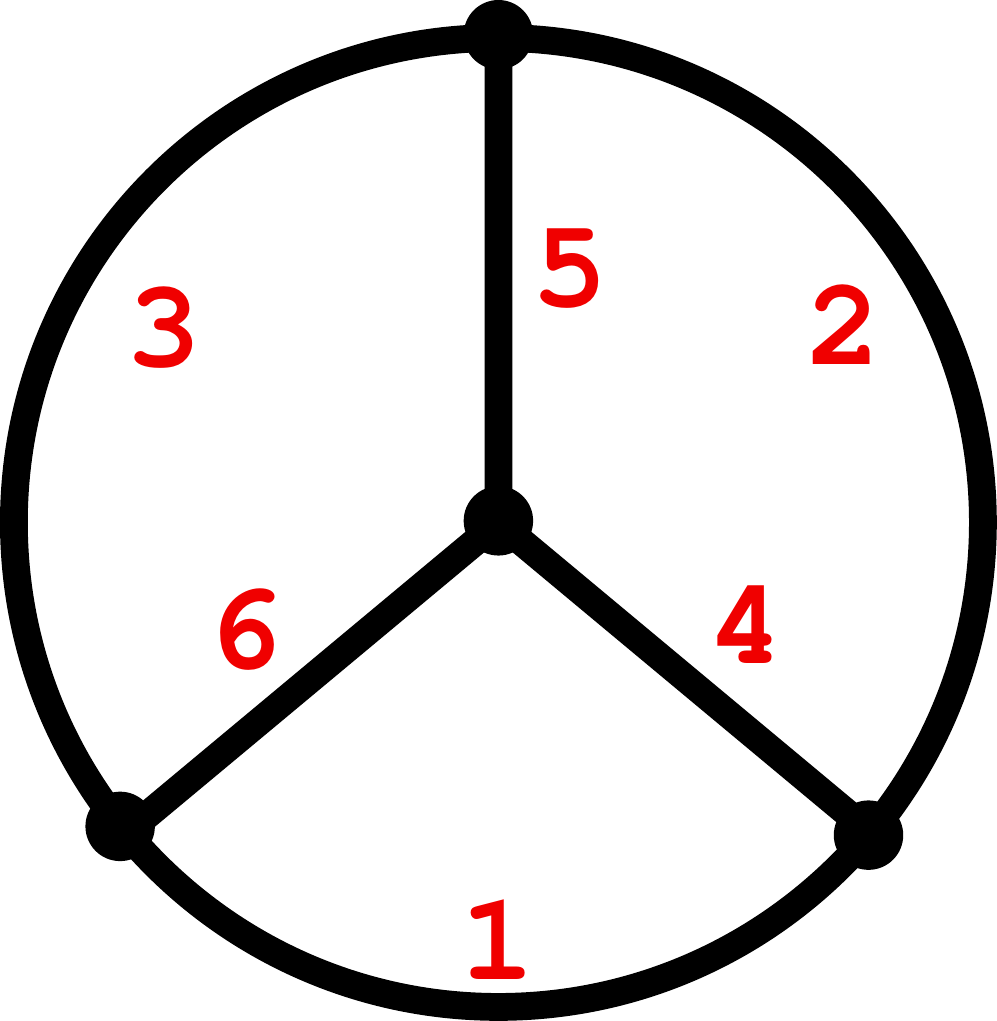}}} & = D_{a_1a_2a_3a_4a_5a_6} =
                                                            \int\frac{d[k_1]d[k_2]d[k_3]}{C_{1;m_1}^{a_1}C_{2;m_2}^{a_2}C_{3;m_3}^{a_3}C_{12;m_4}^{a_4}C_{23;m_5}^{a_5}C_{31;m_6}^{a_6}} \,.
                                                            \intertext{In
                                                            addition
                                                            to
                                                            diagrams
                                                            with six
                                                            lines, we
                                                            also have three-loop digrams with five and four lines,}
                                                            \vcenter{\hbox{\includegraphics[width=2cm]{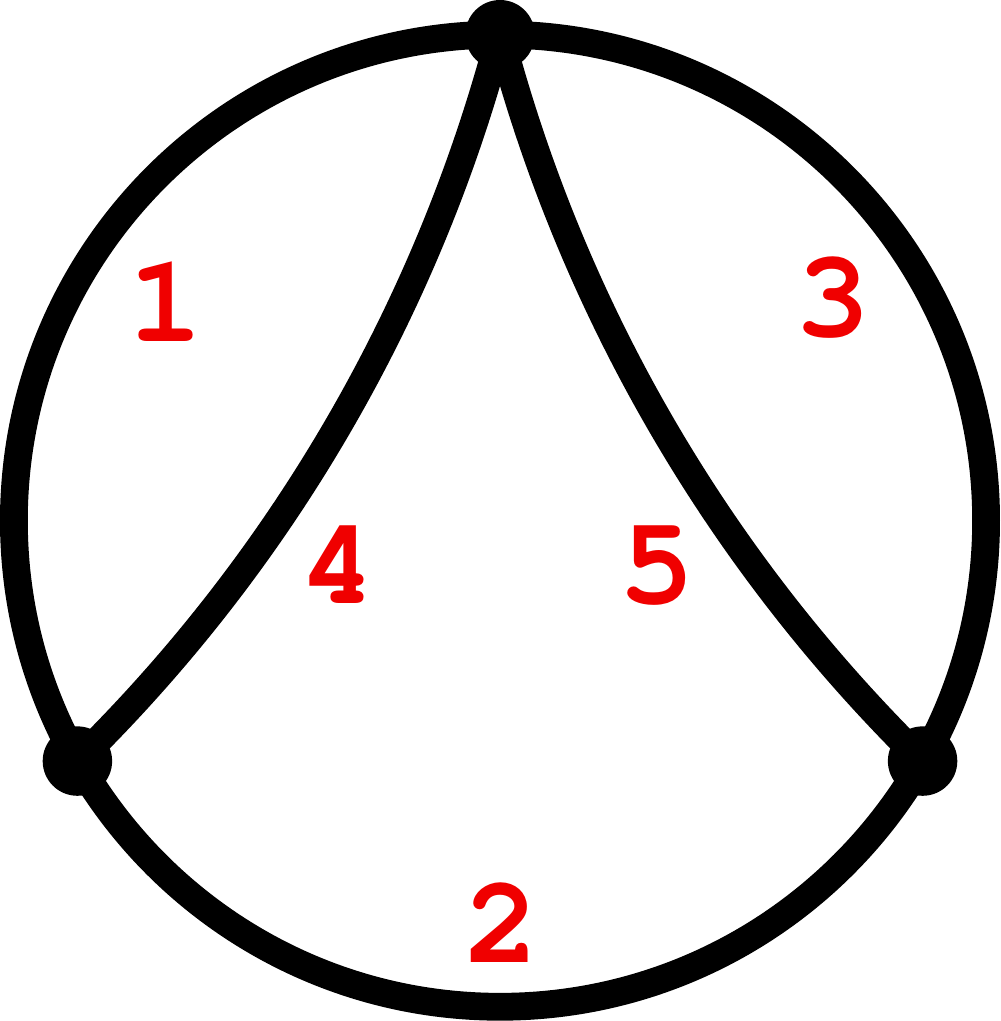}}} & =  E_{a_1a_2a_3a_4a_5} =
                                                                                                                      \int\frac{d[k_1]d[k_2]d[k_3]}{C_{1;m_1}^{a_1}C_{2;m_2}^{a_2}C_{3;m_3}^{a_3}C_{12;m_4}^{a_4}C_{23;m_5}^{a_5}} \,,\\
  \vcenter{\hbox{\includegraphics[width=2cm]{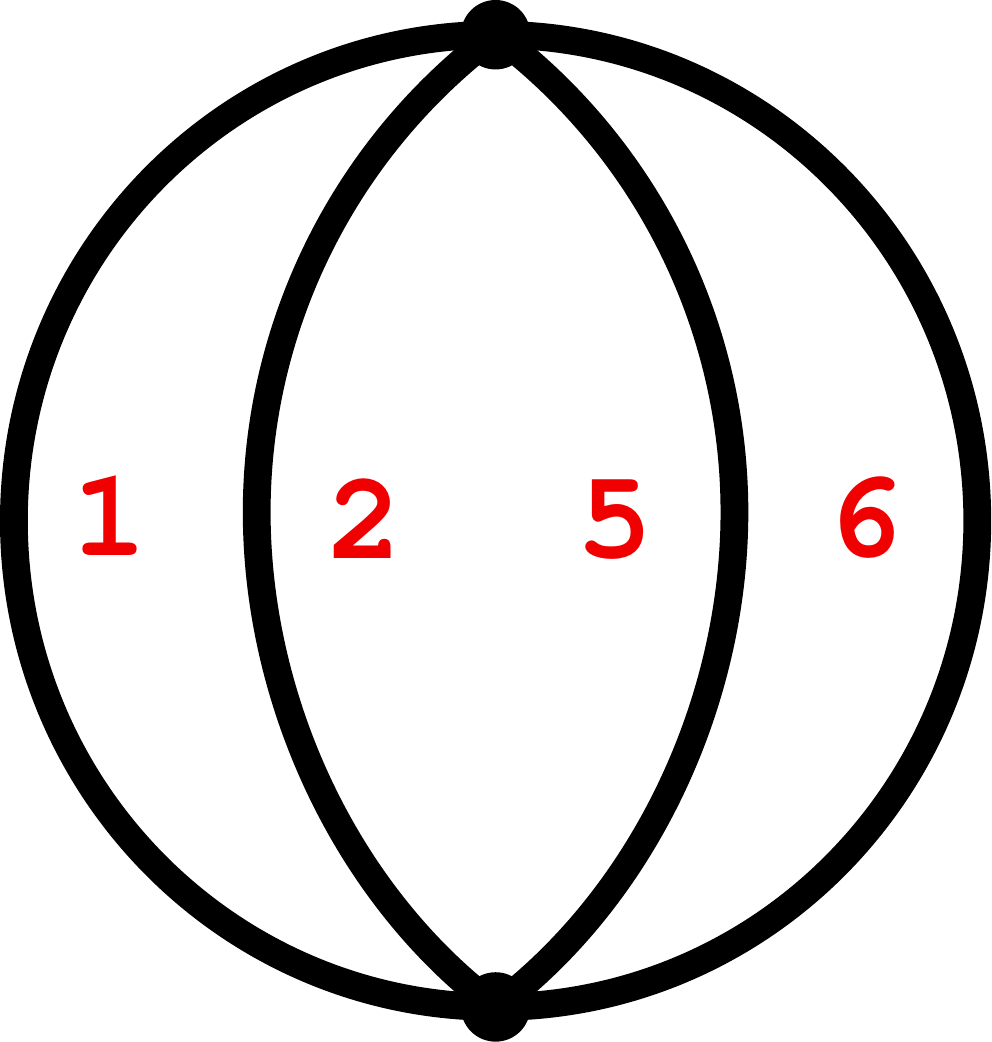}}} & = B_{a_1a_2a_5a_6} =  
                                                            \int\frac{d[k_1]d[k_2]d[k_3]}{C_{1;m_1}^{a_1}C_{2;m_2}^{a_2}C_{23;m_5}^{a_5}C_{31;m_6}^{a_6}} \,,
                                                            \label{eq:fourlines}
                                                            \intertext{as
                                                            well as
                                                            two-loop
                                                            and
                                                            one-loop diagrams,}
                                                            \vcenter{\hbox{\includegraphics[width=2cm]{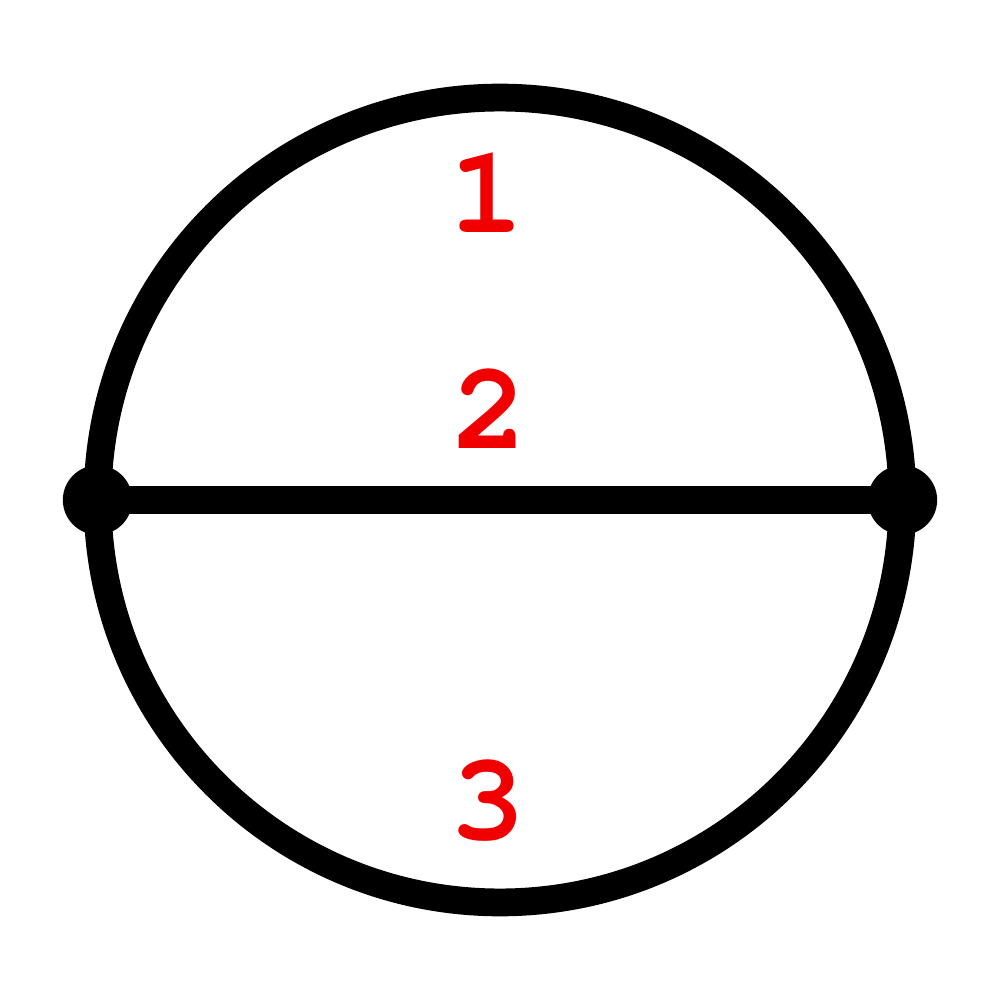}}}
                                                                                                                    & =T_{a_1a_2a_3} =  
                                                                                                                      \int\frac{d[k_1]d[k_2]}{C_{1;m_1}^{a_1}C_{2;m_2}^{a_2}C_{12;m_3}^{a_3}}
                                                                                                                      \,,\quad \vcenter{\hbox{\includegraphics[width=2cm]{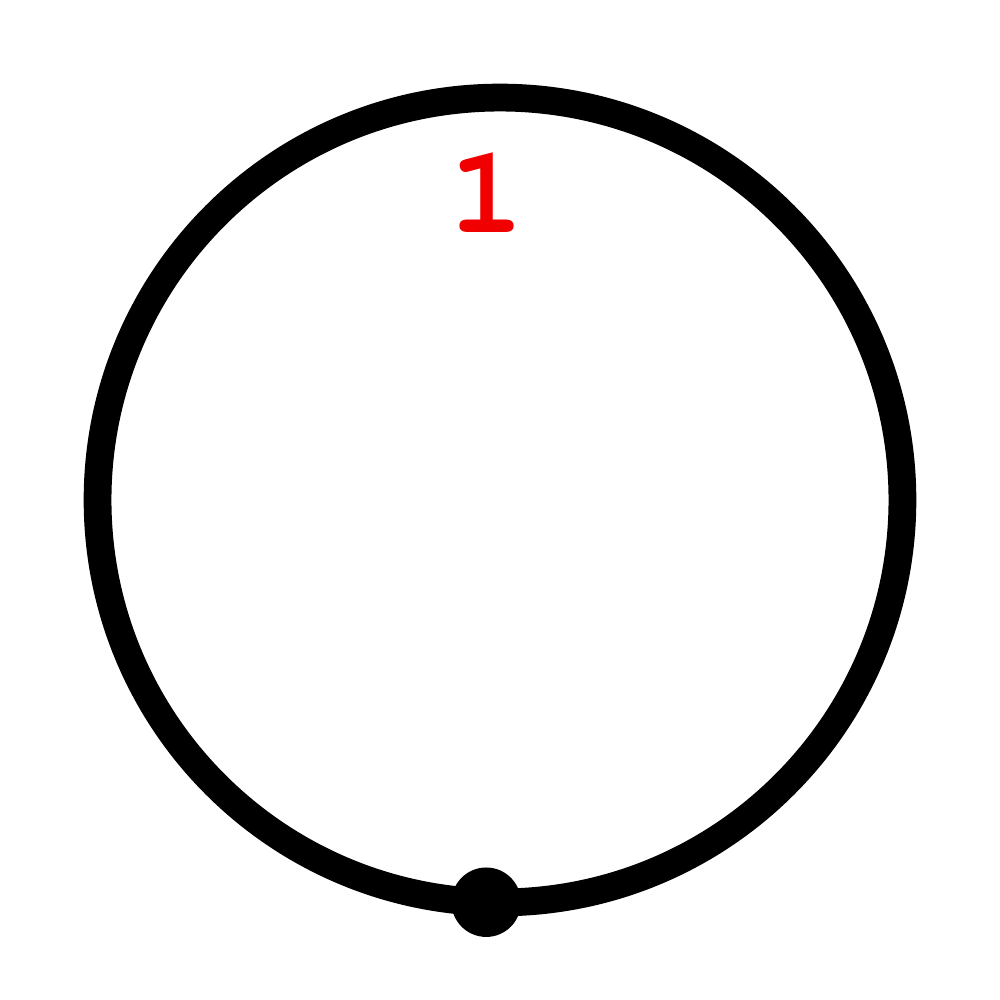}}} = V_{a_1} =
                                                                                                                      \int\frac{d[k_1]}{C_{1;m_1}^{a_1}} \,.
\end{align}


\section{Polylogarithms, algebra of the sixth root of unity, and its subalgebras}

In our study, the key role is played by multiple polylogarithms \cite{Kummer1840,MPL1,Goncharov:1998kja},
defined recursively as repeated integrals,
\begin{align}
  G_{a_1a_2\dots a_w}(z) &= \int\limits_0^{z} \frac{dt_1}{t_1-a_1}
                           \underbrace{
                           \int\limits_0^{t_1} \frac{dt_1}{t_2-a_2} \dots
                           \int\limits_0^{t_{w-1}} \frac{dt_w}{t_w-a_w}
                           }_{G_{a_2\dots a_w}(t_1)}
                           \,, \qquad a_w\neq0 \,,
                           \label{eq:Gdef1}
\end{align}
where $a_1,a_2,\dots,a_w$ and $z$ are complex numbers. The definition in eq.~(\ref{eq:Gdef1}) is modified
in the case of $q$ trailing zero indices in the following way:
\begin{align}
  G_{a_1a_2\dots a_{w-q}00\dots0}(z) &= \int\limits_0^{z} \frac{dt_1}{t_1-a_1}
                                       \int\limits_0^{t_1} \frac{dt_1}{t_2-a_2} \dots
                                       \int\limits_0^{t_{w-q-1}} \frac{dt_{w-q}}{t_{w-q}-a_{w-q}} \, \frac{\ln^q(t_{w-q})}{q!} \,.
                                       \label{eq:Gdef2}
\end{align}
The integer number $w$ is called the {\it weight} of the polylogarithm.

The functions $G$ obey the so-called shuffle and stuffle relations. In particular,
any product of two $G$ functions with the same argument and weights $w_1$ and $w_2$ can be
rewritten as a linear combination of $G$ functions of weight $w_1+w_2$. In other words,
polylogarithms form a graded algebra. 

The algebra of the sixth root of unity ${\cal A}_\omega$ 
is obtained from general polylogarithms by
restricting all $a_j$ to the seven-letter alphabet $\{0,\omega^0,\omega^1,\omega^2\dots,\omega^5\}$, 
where 
\begin{align}
  \omega=\exp\Big(\frac{i\pi}{3}\Big) 
  \label{eq:omega}
\end{align}
is a primitive sixth root of unity. At arbitrary argument $z$,
such functions include the so-called inverse-binomial-sums
functions~\cite{Kalmykov:2000qe,Davydychev:2003mv,Fleischer:1998nb,Weinzierl:2004bn,Kalmykov:2010xv,Ablinger:2014bra}
and are related to cyclotomic polylogarithms \cite{Ablinger:2011te}.
At $z=1$, they represent a set of irrational\footnote{This means irrational up to the precision used for the \texttt{PSLQ} reconstruction.} constants, which is relevant for the description of some single-scale massive diagrams (in particular, three-loop vacuum bubble and two-loop on-shell
self-energy diagrams). The complete basis of the algebra of the sixth root of unity ${\cal A}_\omega$ through
weight 6 has recently been constructed in ref.~\cite{Henn:2015sem}.

In this work, we construct the basis of the subalgebra of ${\cal A}_\omega$ formed by the 
harmonic polylogarithms~\cite{Remiddi:1999ew} $H_{n_1\dots n_p}(z)$ of arguments $z_k=\omega^k$. 
We shall call such an algebra ${\cal A}_{H(\omega^k)}$.
The harmonic polylogarithms
are defined similarly to~eqs.~(\ref{eq:Gdef1})--(\ref{eq:Gdef2}), but now the parameters $a_j$ can only take
the values $-1,0,+1$. For historical reasons, there is also a difference in the overall sign. 
Specifically, the harmonic-polylogarithm functions $H_{n_1\dots n_p}(z)$ are related to the
generalized polylogarithms $G$ via
\begin{align}
  H_{n_1n_2\dots n_w}(z) &=   (-1)^{\displaystyle(\mbox{number of }n_j=1)} G_{n_1n_2\dots n_w}(z) \,,
                           \label{eq:Hdef}
\end{align}
where $n_j=-1,0,+1$.

Using the scaling properties of the polylogarithms together with the shuffle relations,
it is easy to show that any element of the form $H_{n_1 \dots n_p}(\omega^k)$ can be
rewritten as $G_{\omega^{k_1}\dots\omega^{k_p}}(1)$ with some integers $k_j=0,1,\ldots,5$.
For example, we have $H_{1,0,-1}(\omega)=G_{1/\omega,0,-1/\omega}(1)=G_{\omega^5,0,\omega^2}(1)$.
In other words,
\begin{align}
  {\cal A}_{H(\omega^k)} \subset {\cal A}_\omega\,,  \qquad k=0,1,\dots,5\,.
\end{align}

In particular, ${\cal A}_{H(\omega^0)}\equiv{\cal A}_{H(1)}$  and
${\cal A}_{H(\omega^3)}\equiv{\cal A}_{H(-1)}$ form the shuffle
algebra of the multiple zeta values (see, e.g.,\ ref.~\cite{MZV}).
The pairs (${\cal A}_{H(\omega^1)}$, ${\cal A}_{H(\omega^5)}$) 
and (${\cal A}_{H(\omega^2)}$, ${\cal A}_{H(\omega^4)}$) are related to each other by complex
conjugation, since $\overline{\omega^1}=\omega^5$ and $\overline{\omega^2}=\omega^4$. 
Moreover, all ${\cal A}_{H(\omega^k)}$, $k=1,2,4,5$ are isomorph. We prove this by explicitly constructing
the relation between the corresponding bases over $\mathbb{Q}$. Thus, we need to
consider only ${\cal A}_{H(\omega)}$. We can split each element into its real and imaginary parts,
\begin{align}
  H_{n_1 \dots n_w}(\omega) = \Re H_{n_1 \dots n_w}(\omega) + i \Im
  H_{n_1 \dots n_w}(\omega) \,.
\end{align}
Note that $\Re {\cal A}_{H(\omega)}$ can involve products of even numbers of elements
from $\Im {\cal A}_{H(\omega)}$, while $\Im {\cal A}_{H(\omega)}$ can involve products 
of elements from $\Re {\cal A}_{H(\omega)}$.

Using the shuffle relations and the PSLQ algorithm~\cite{PSLQ}, we construct the real and imaginary
bases for the weights $w=1,\dots,6$.
The numbers of the basis elements at each weight for the algebra of the sixth root of unity ${\cal A}_\omega$
and for ${\cal A}_{H(\omega)}$ are summarized in
table~\ref{tab:one}.
\begin{table}
  \begin{center}
    \begin{tabular}{ | c | c | c | c | c |}
      \hline\hline
      $w$ & $\Re {\cal A}_\omega$ & $\Im {\cal A}_\omega$ 
      & $\Re {\cal A}_{H(\omega)}$ & $\Im {\cal A}_{H(\omega)}$ \\
      \hline\hline
      1 & 2   & 1   & 1   & 1   \\ \hline
      2 & 5   & 3   & 3   & 3   \\ \hline
      3 & 12  & 9   & 8   & 8   \\ \hline
      4 & 30  & 25  & 21  & 21  \\ \hline
      5 & 76  & 68  & 55  & 55  \\ \hline
      6 & 195 & 182 & 144 & 144 \\ 
      \hline\hline
    \end{tabular}
    \caption{\label{tab:one}%
      Values of $\Re {\cal A}_\omega$, $\Im {\cal A}_\omega$,
      $\Re {\cal A}_{H(\omega)}$, and $\Im {\cal A}_{H(\omega)}$ for $w=1,\ldots,6$.} 
  \end{center}
\end{table}
The results for $\Re{\cal A}_\omega$ and $\Im{\cal A}_\omega$ were obtained
by explicit calculation in ref.~\cite{Henn:2015sem}.
The results for $\Re {\cal A}_{H(\omega)}$ and $\Im {\cal A}_{H(\omega)}$
are obtained in this work.
Unlike ${\cal A}_\omega$, the real and imaginary parts of ${\cal A}_{H(\omega)}$
have the same numbers of basis elements. The integer sequence $1,3,8,21,55,144,\dots$
corresponds to $\{F_{2w}\}$, $w=1,2,3,4,5,6,\dots$, where $F_{j}$ denotes the $j$-th 
Fibonacci number.

We shall denote the uniform bases of $\Re{\cal A}_{H(\omega)}$ and $\Im{\cal A}_{H(\omega)}$
for fixed weight $w$ as $\Re H_w$ and $\Im H_w$, respectively.
In the next sections, we apply the constructed bases $\Re H_w$ and $\Im H_w$
to the evaluation of the three-loop massive vacuum bubble diagrams.


\section{Evaluation of the three-loop vacuum bubble integrals}

\begin{figure}[h]
  \captionsetup[subfigure]{labelformat=empty}
  \centering
  \begin{subfigure}{2cm}
    \centering
    \includegraphics[width=\textwidth]{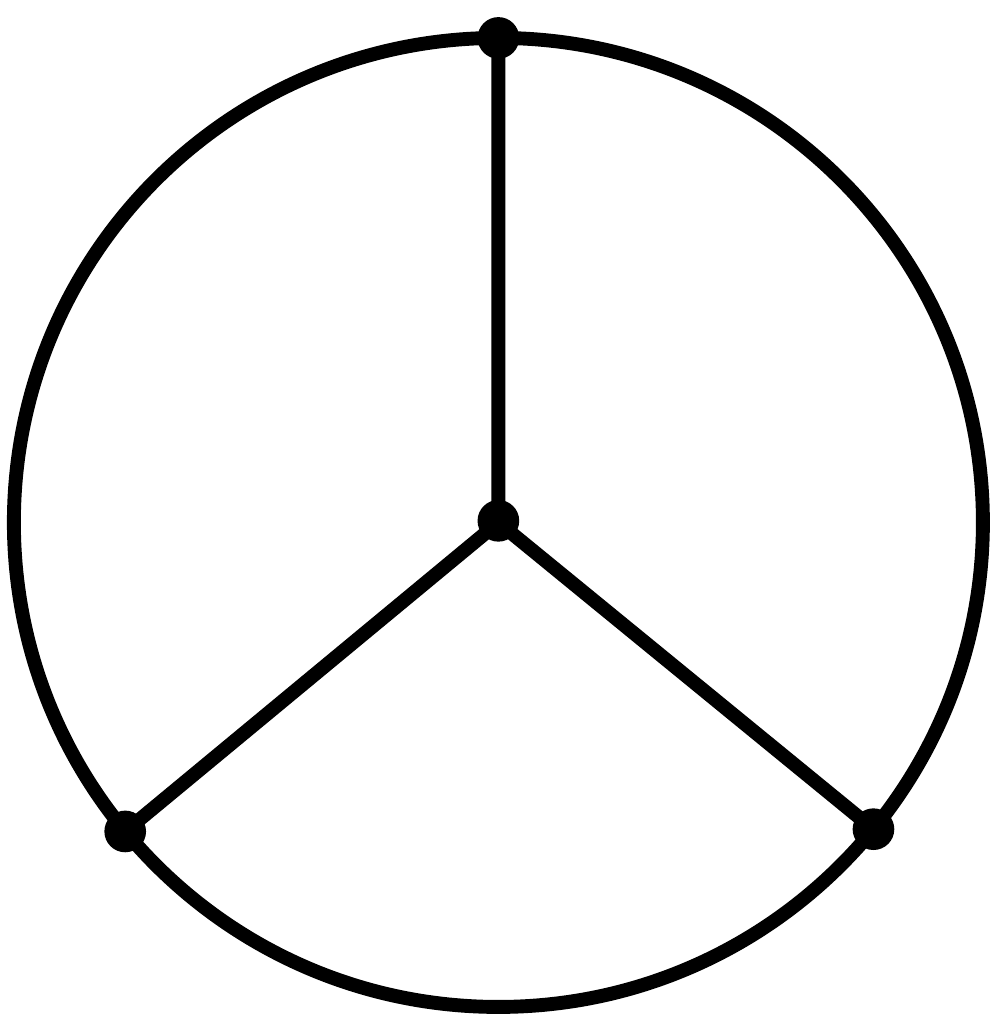}
    \caption{$\bf D_6$}
  \end{subfigure}%
  ~ 
  \begin{subfigure}{2cm}
    \centering
    \includegraphics[width=\textwidth]{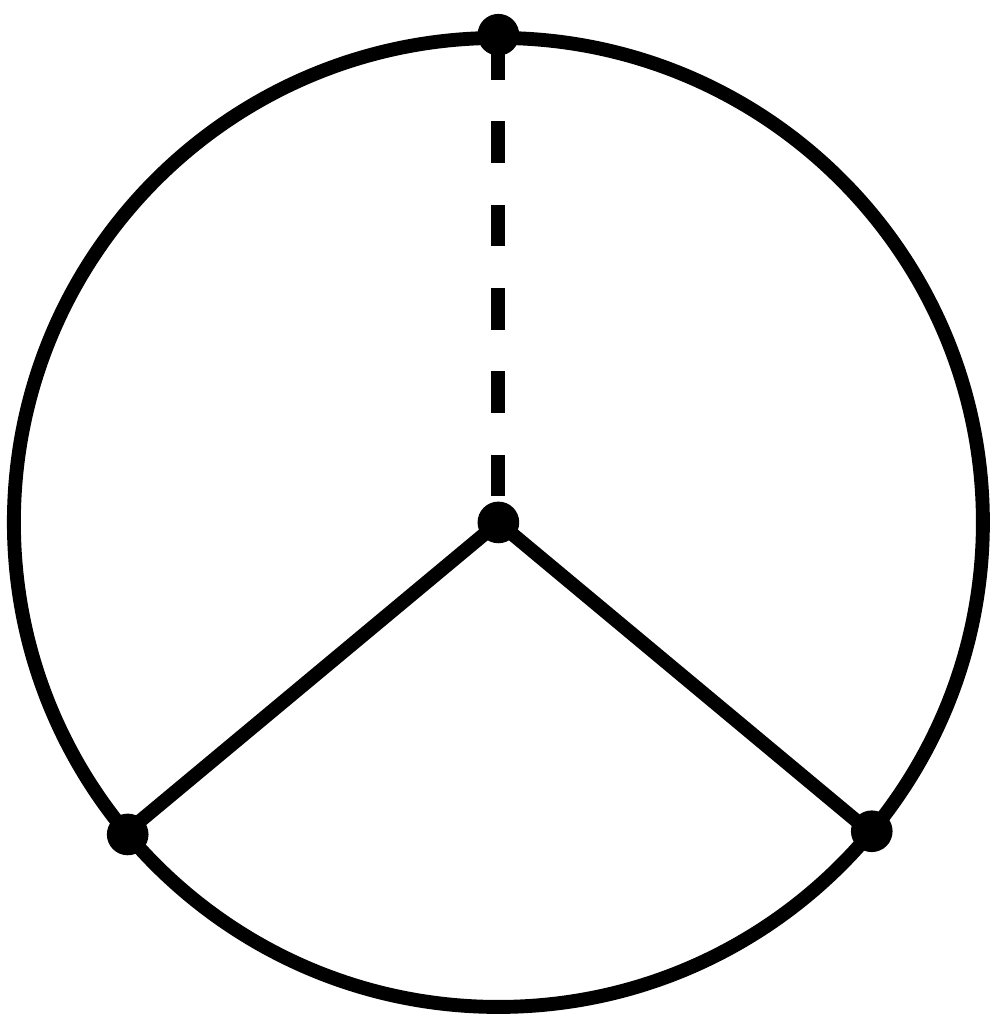}
    \caption{$\bf D_5$}
  \end{subfigure}
  ~ 
  \begin{subfigure}{2cm}
    \centering
    \includegraphics[width=\textwidth]{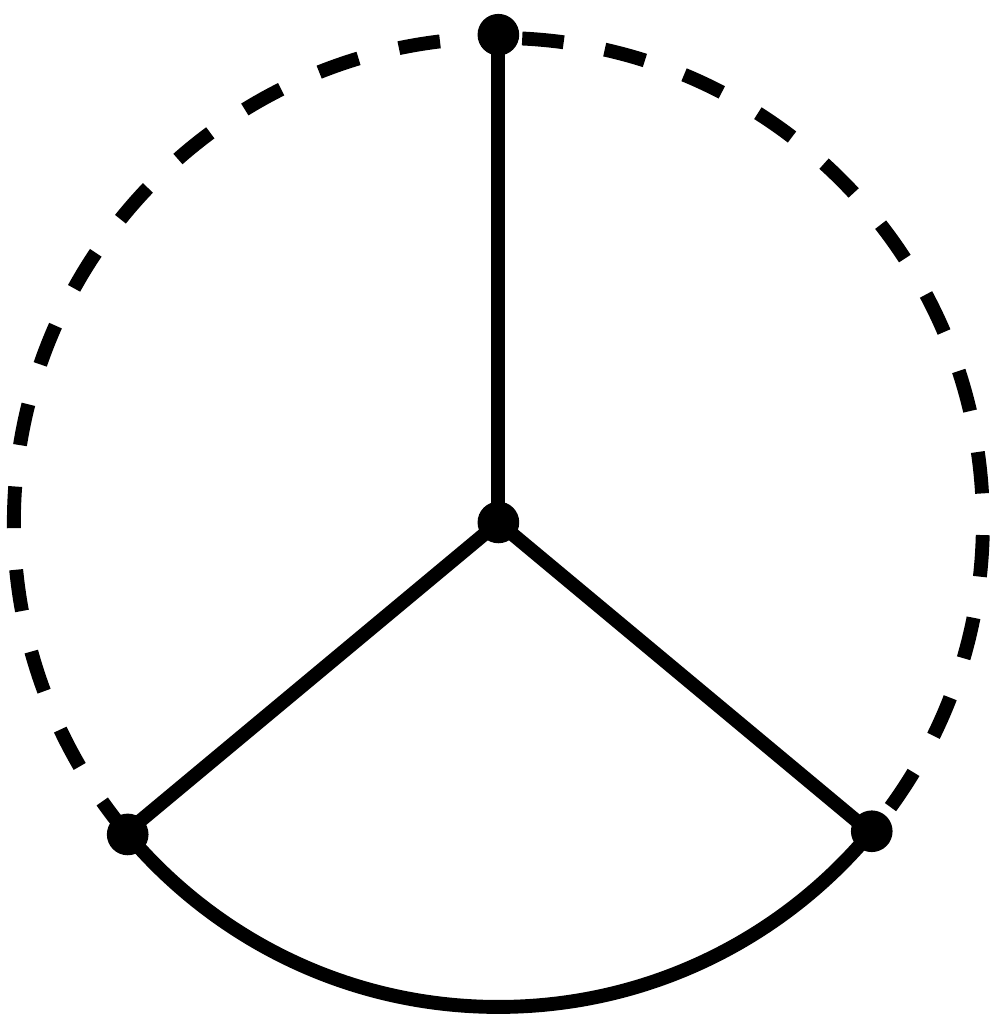}
    \caption{$\bf D_4$}
  \end{subfigure}
  ~ 
  \begin{subfigure}{2cm}
    \centering
    \includegraphics[width=\textwidth]{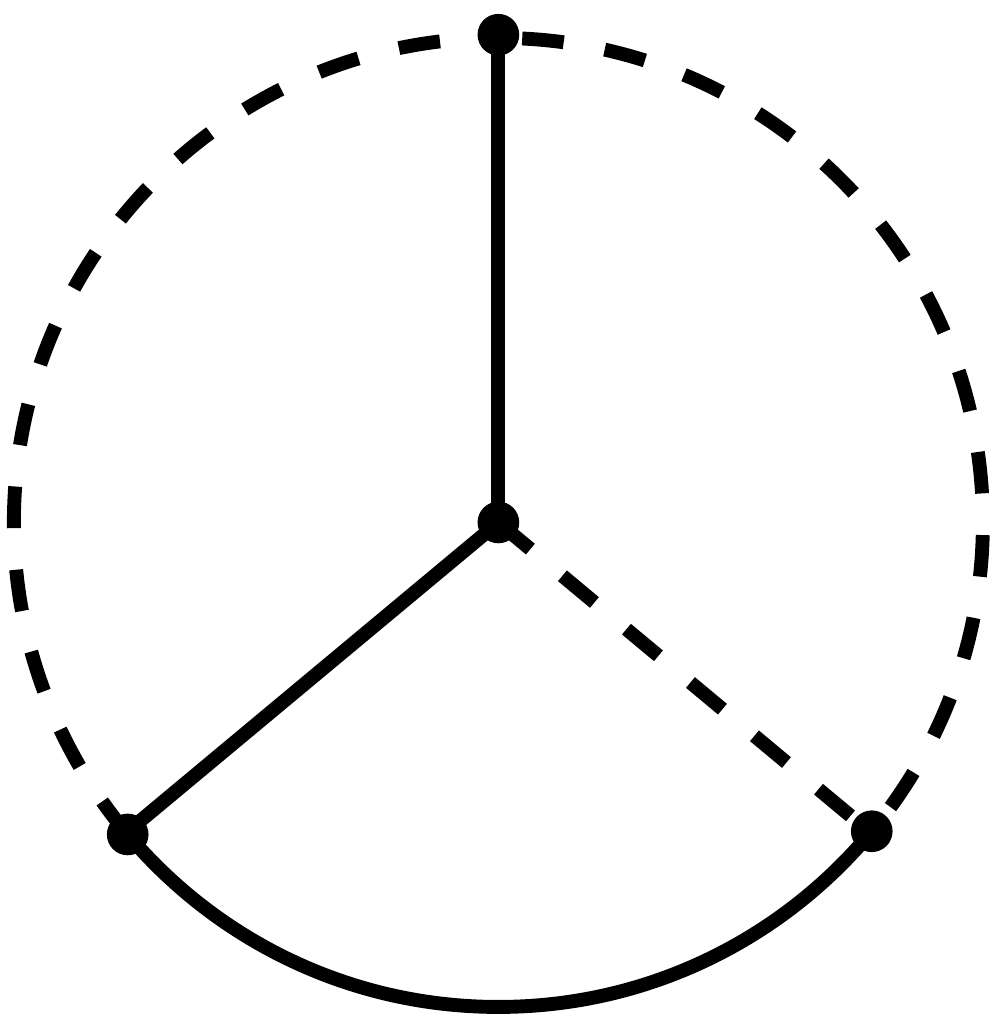}
    \caption{$\bf D_3$}
  \end{subfigure}
  ~ 
  \begin{subfigure}{2cm}
    \centering
    \includegraphics[width=\textwidth]{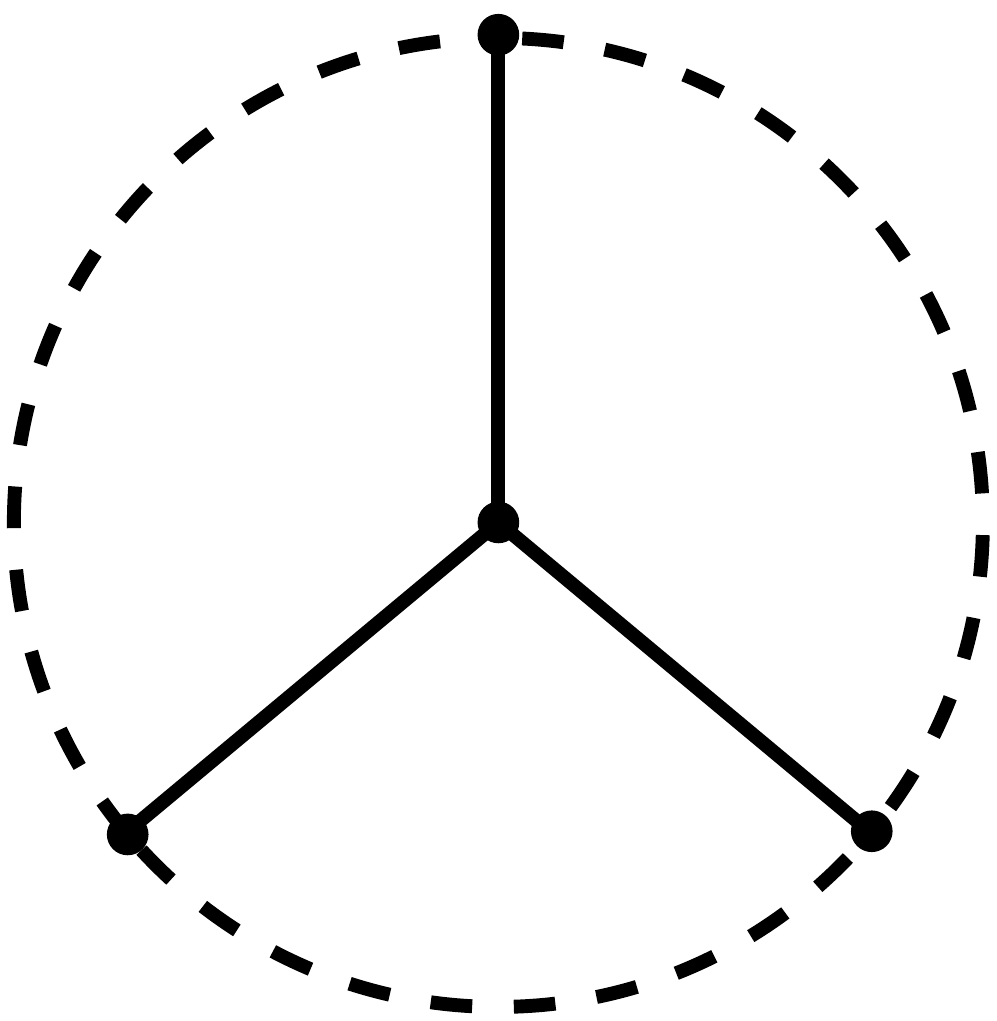}
    \caption{$\bf D_M$}
  \end{subfigure}
  ~ 
  \begin{subfigure}{2cm}
    \centering
    \includegraphics[width=\textwidth]{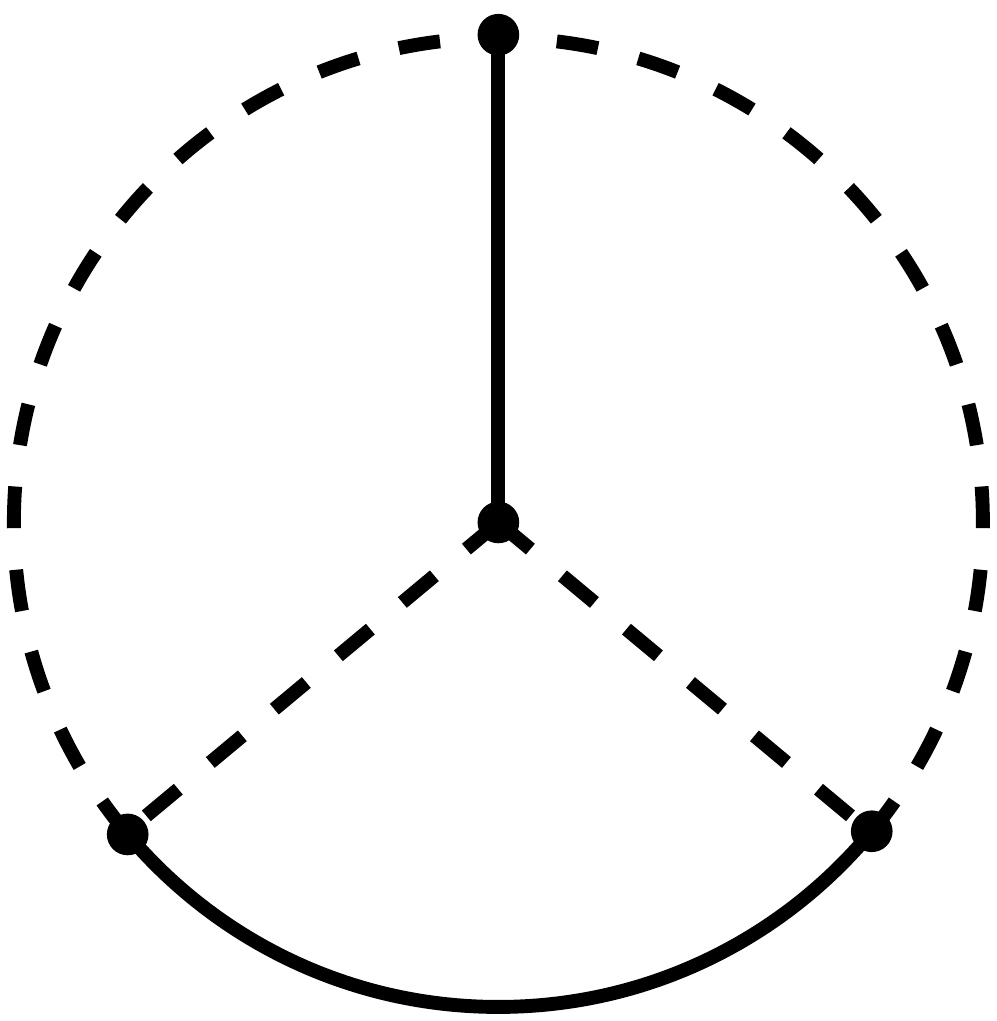}
    \caption{$\bf D_N$}
  \end{subfigure}
  \\[10mm]
  \begin{subfigure}{2cm}
    \centering
    \includegraphics[width=\textwidth]{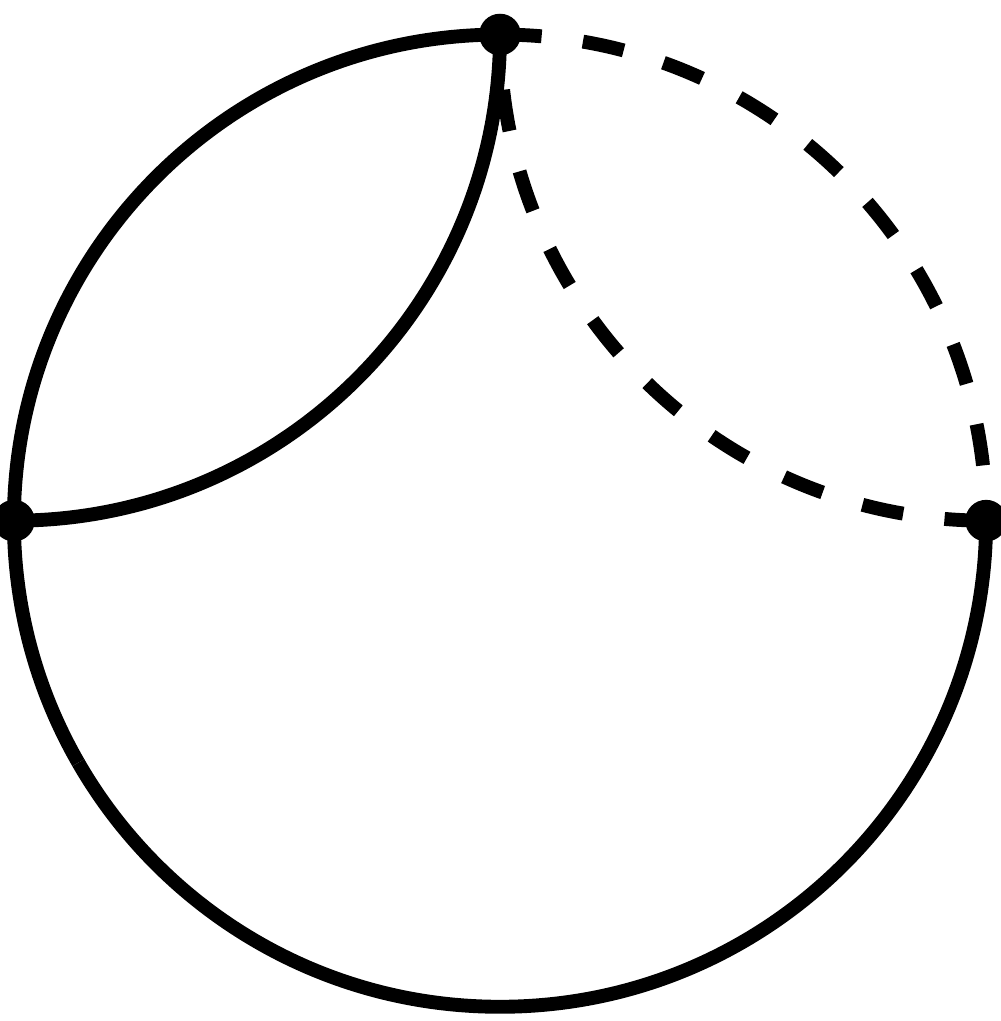}
    \caption{$\bf E_3$}
  \end{subfigure}
  ~ 
  \begin{subfigure}{2cm}
    \centering
    \includegraphics[width=\textwidth]{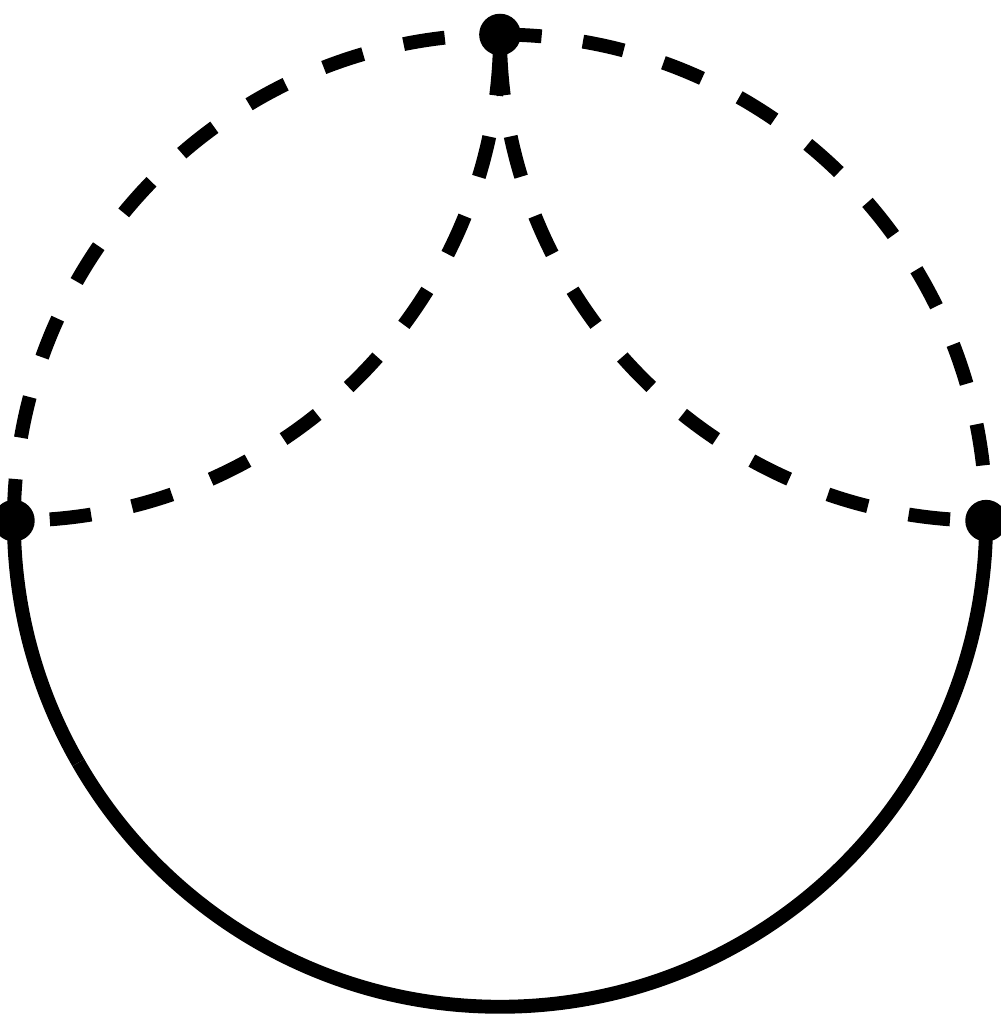}
    \caption{$\bf E_1$}
  \end{subfigure}
  ~
  \begin{subfigure}{2cm}
    \centering
    \includegraphics[width=\textwidth]{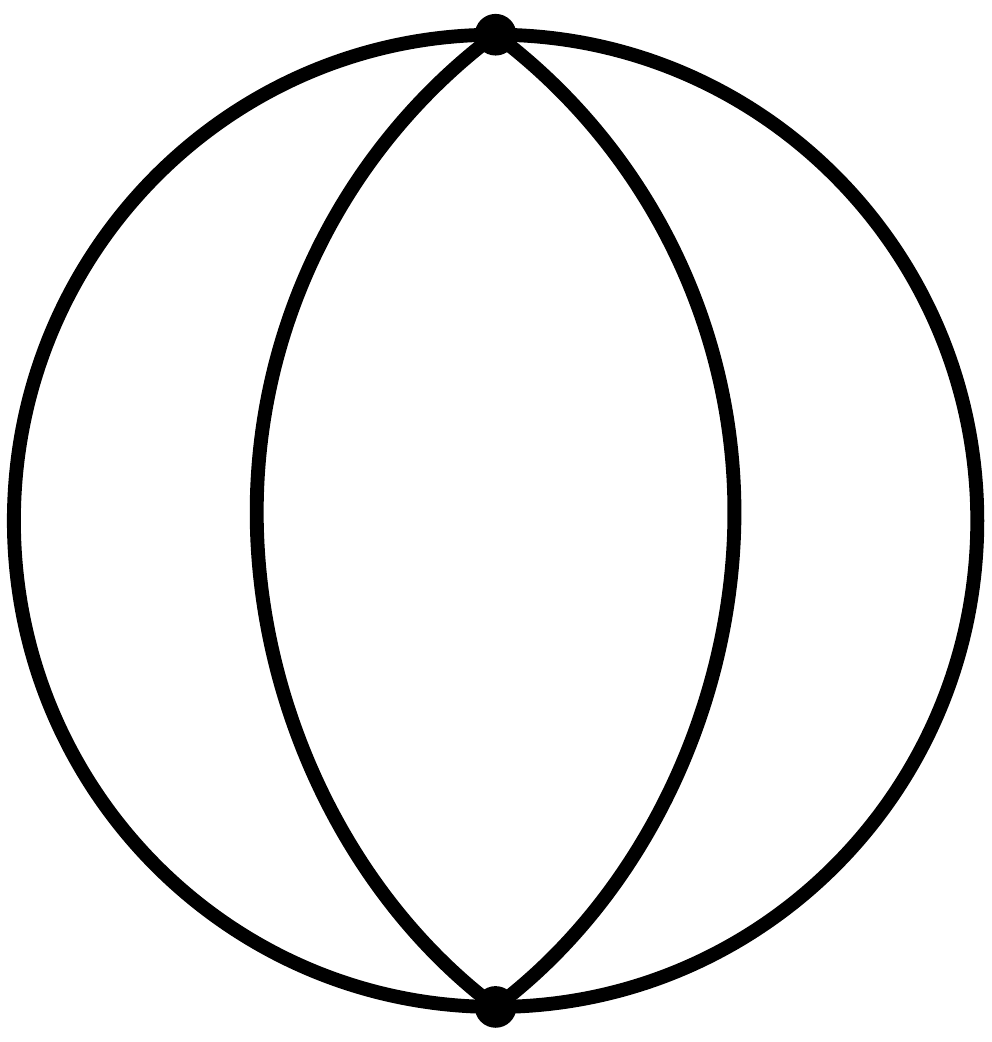}
    \caption{$\bf BN$}
  \end{subfigure}
  ~ 
  \begin{subfigure}{2cm}
    \centering
    \includegraphics[width=\textwidth]{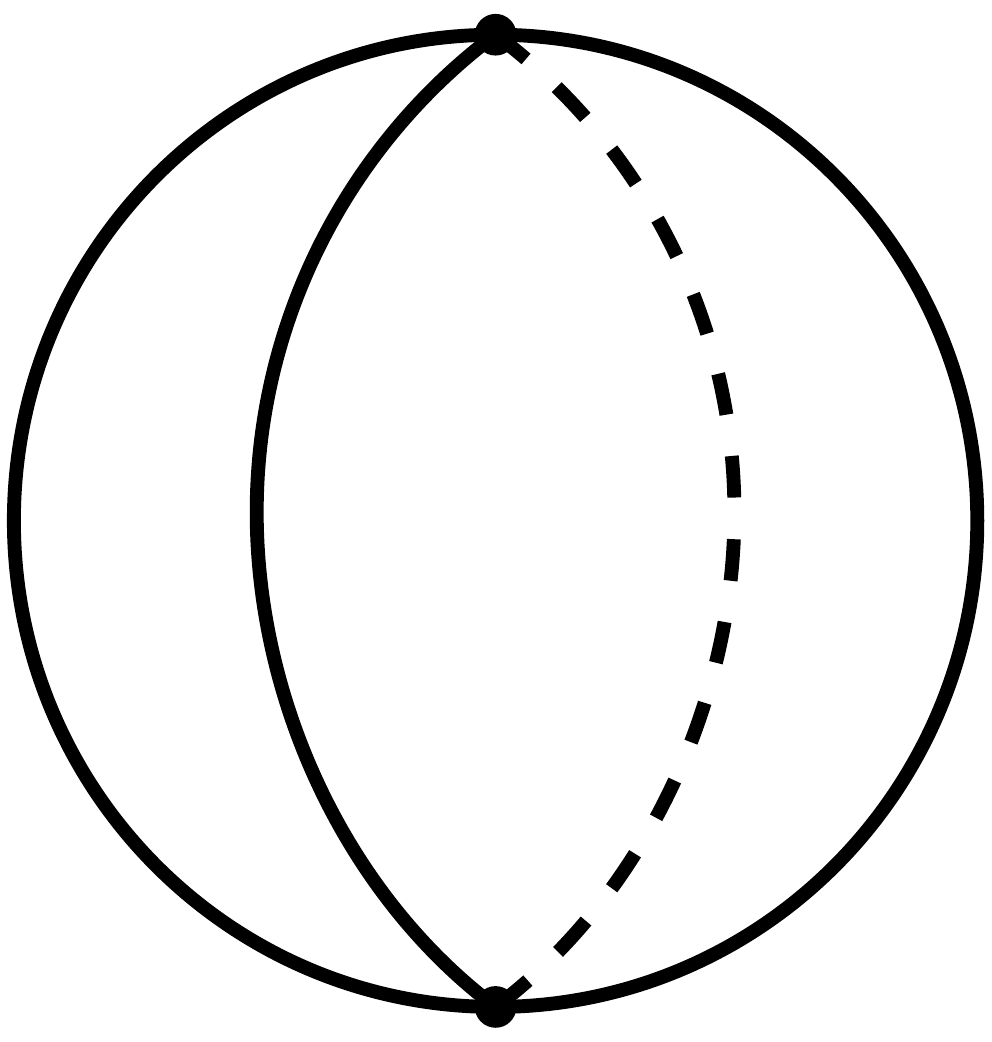}
    \caption{$\bf BN_1$}
  \end{subfigure}
  ~ 
  \begin{subfigure}{2cm}
    \centering
    \includegraphics[width=\textwidth]{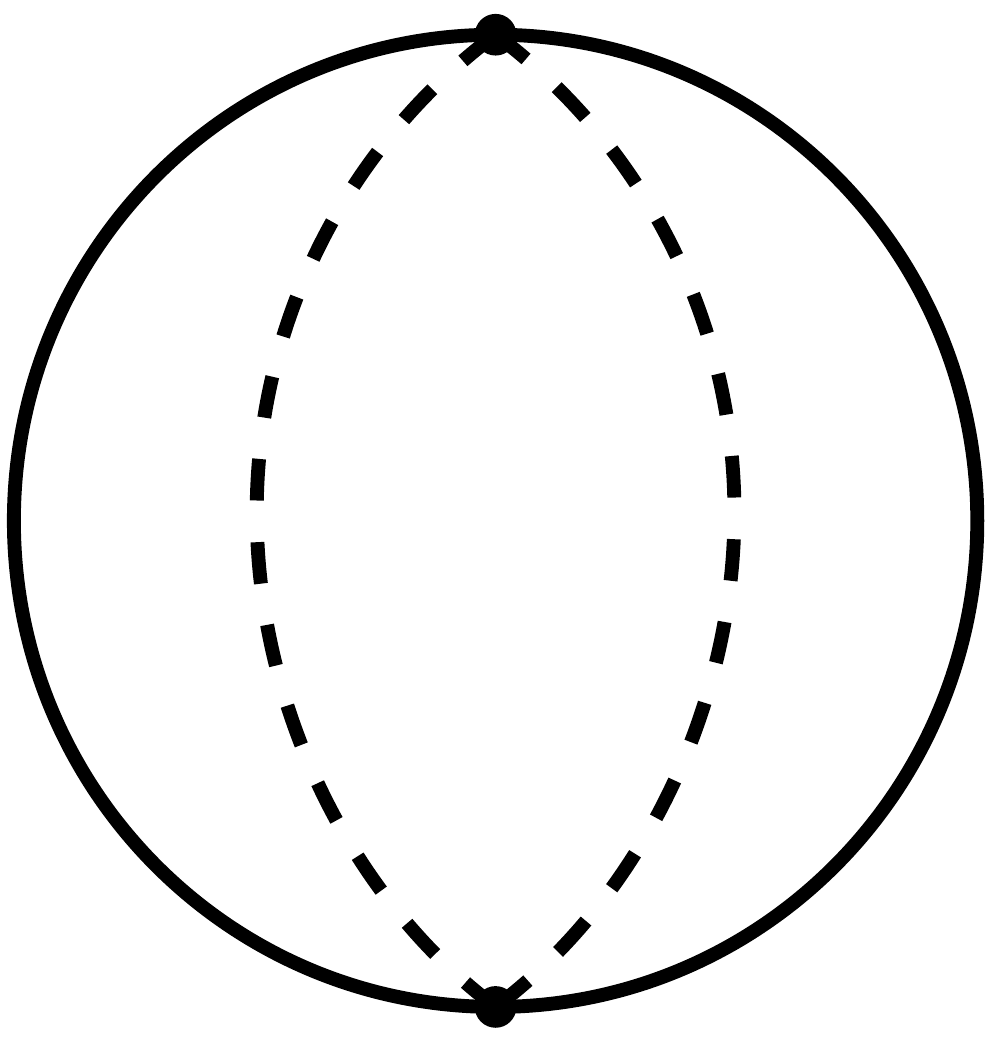}
    \caption{$\bf BN_2$}
  \end{subfigure}
  ~ 
  \begin{subfigure}{2cm}
    \centering
    \includegraphics[width=\textwidth]{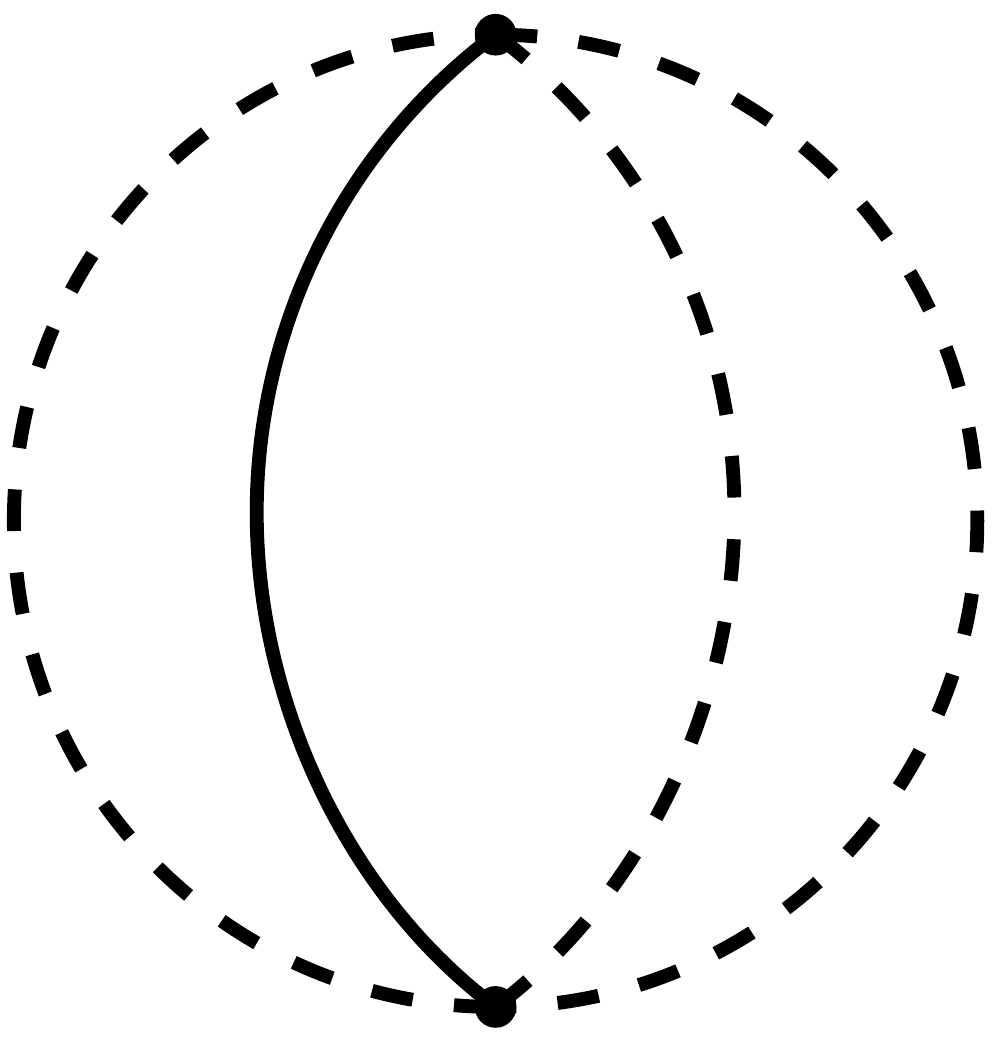}
    \caption{$\bf BN_3$}
  \end{subfigure}
  \caption{Full set of non-factorizing master integrals.
    Solid and dashed lines correspond to massive and massless
    scalar propagators, respectively.}
  \label{fig:tad1234}
\end{figure}

\begin{figure}[h]
  \captionsetup[subfigure]{labelformat=empty}
  \centering
  \begin{subfigure}{2cm}
    \centering
    \includegraphics[width=\textwidth]{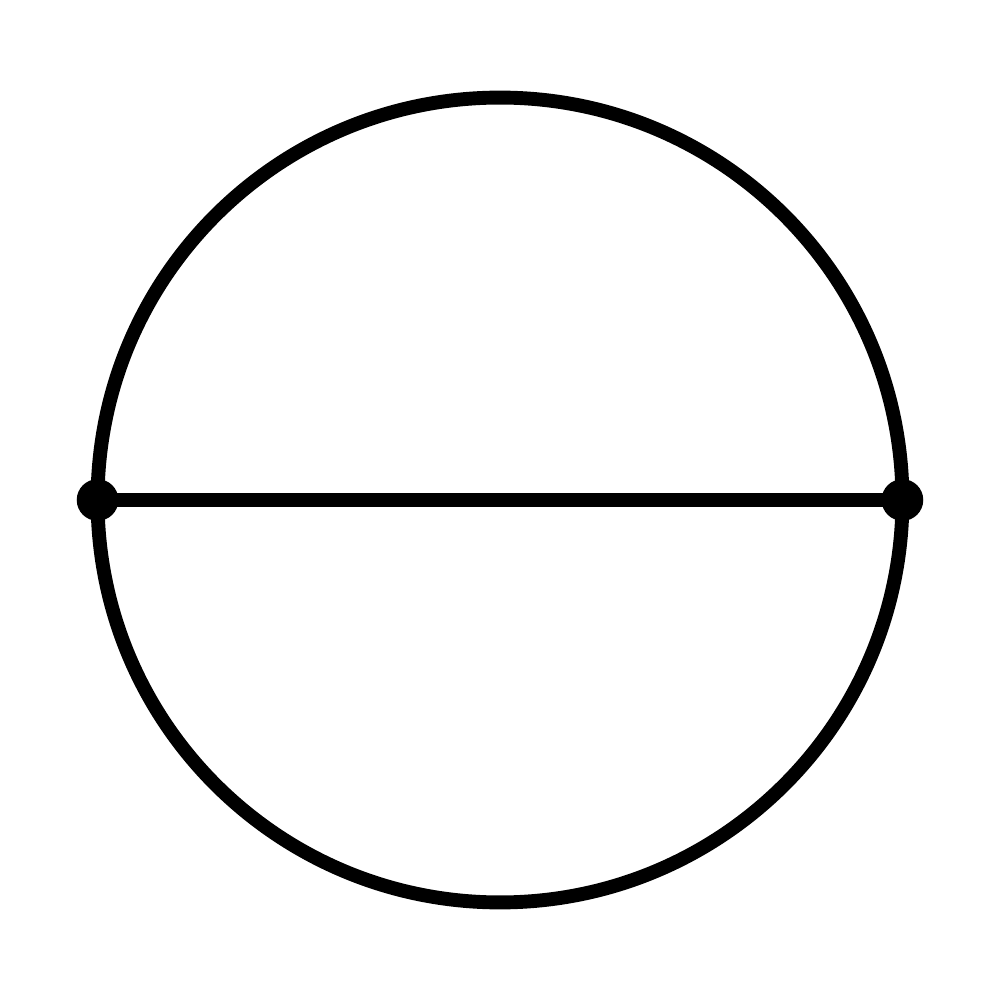}
    \caption{$\bf T_{111}$}
  \end{subfigure}
  ~ 
  \begin{subfigure}{2cm}
    \centering
    \includegraphics[width=\textwidth]{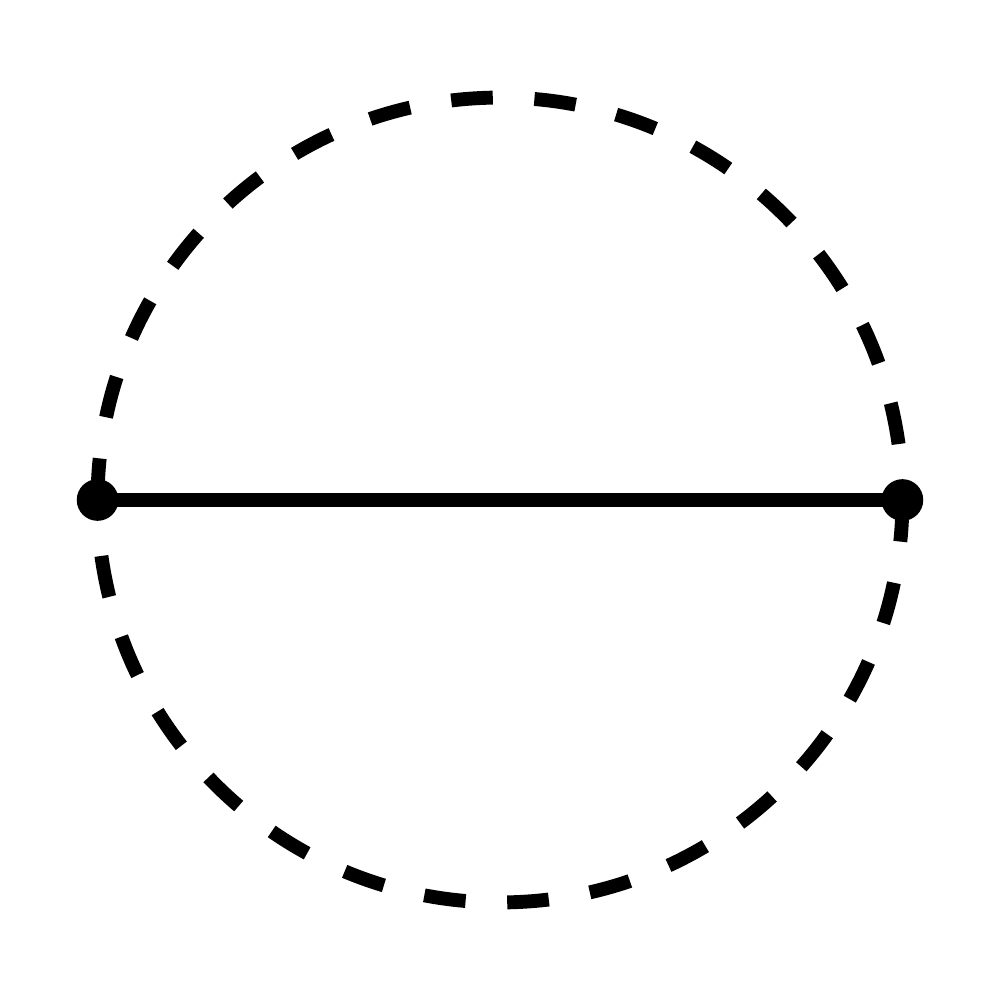}
    \caption{$\bf T_{100}$}
  \end{subfigure}
  ~ 
  \begin{subfigure}{2cm}
    \centering
    \includegraphics[width=\textwidth]{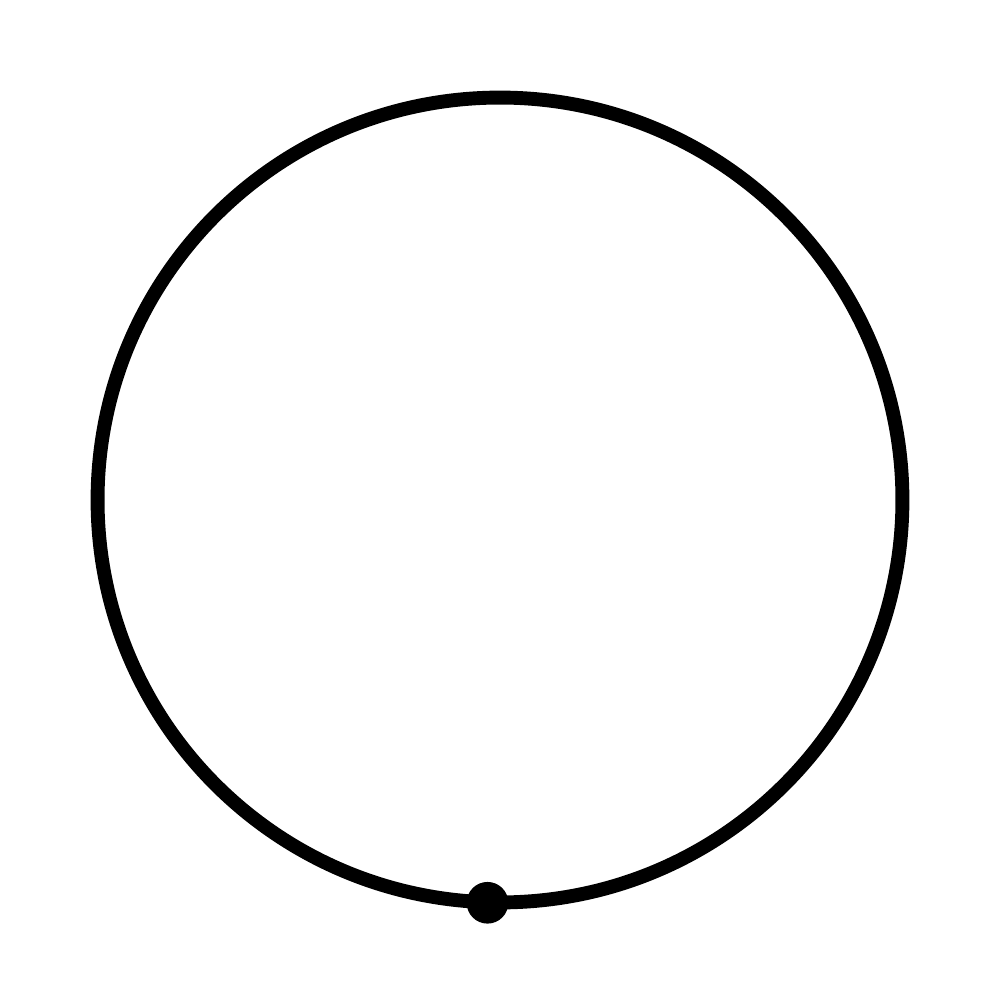}
    \caption{$\bf V_1$}
  \end{subfigure}
  \caption{Two-loop and one-loop master integrals. The line coding is the same as in figure~\ref{fig:tad1234}.}
  \label{fig:tad12}
\end{figure}

Using integration-by-parts relations, it is possible to reduce any three-loop bubble integral
with a single scale to a set of twelve three-loop master integrals, two two-loop integrals,
and one one-loop bubble. These diagrams are shown in figures~\ref{fig:tad1234} and
\ref{fig:tad12}.

It is the goal of this paper to evaluate these master integrals analytically in terms
of polylogarithms through weight six.

In the previous section, we discussed the construction of the bases of
$\Re{\cal A}_{H(\omega)}$
and $\Im{\cal A}_{H(\omega)}$. We now use these bases to reconstruct the analytic expressions
for the $\eps$ expansions of these diagrams using the PSLQ algorithm~\cite{PSLQ}.
For that purpose, we first need
a precise numerical value of each diagram.

Specifically, for the fully massive diagrams $\bf D_6$ and $\bf E_5$, we make use of the series obtained 
with the help of the DRA method, based on dimensional recurrence relations and
analyticity, presented in ref.~\cite{Lee:2012hp} and summed with the
help of the \texttt{SummerTime} package~\cite{Lee:2015eva}. Within a
few hours, we were able to get 20,000 decimal figures of precision for
these diagrams.

The general method of calculation which is used in this work and is applicable to all the considered diagrams 
consists in writing the systems of differential equations
for the integrals and solving them later by the Frobenius method.
In the first step, instead of a single-scale diagram with mass $m$, we introduce a similar diagram
with two different masses, $m_1$ and $m_2$. Then, using integration by parts, 
we can write the system of differential equations in the mass ratio $z=m_1^2/m_2^2$. In general, these
equations cannot be solved analytically. We solve them as series of the form $\sum_n z^n c_n$.
The unknown coefficients $c_n$ are to be determined by substituting the series in the differential equations.

We also need the boundary conditions. The easiest choice in our case is the boundary conditions
at $z=0$, which correspond to a single-scale bubble integral with a smaller number of massive lines.

Finally, we set $z=1$ in the series solution in order to recover the original diagram. The summation
of the series is done numerically. 
In this way, we are able to evaluate integrals to an accuracy
of typically 4,000 to 10,000 decimal figures depending on the diagram. 
For that purpose, we need to sum up to 20,000 terms in the $n$ sum in some cases.

Let us consider as an example the integral $\bf D_N$. There are two massive lines in this diagram.
Instead of two equal masses, we introduce now two different masses, one of which we set to unity.
Thus, we set in eq.~(\ref{eq:int-def}) $m_1=z$, $m_5=1$, and $m_2=m_3=m_4=m_6=0$.
With such masses, we have the following set of master integrals, which depend on $z$:
\begin{align}
  D_{100111}, D_{101110}, D_{201110}, D_{101011}, D_{011111}, D_{111110}, D_{111111} \,,
  \label{eq:DN_MI}
\end{align}
where we use the definition in eq.~(\ref{eq:int-def}).

Let us denote, for brevity, the integrals in eq.~(\ref{eq:DN_MI}) as $f_1,\dots,f_7$ in this very order.
Then, the functions $f_1(z),\dots,f_7(z)$ obey a system of
linear differential equations in the variable $z$, which reads:
\begin{align}
  &  z^2 (1 + z) f'_7 + \frac{1}{2} (d-4) z (1 + 2 z) f_7 =
    \nonumber\\
  & \qquad
    (d-3) z f_6 + (d-3) z f_5 + (d-2) z f_4 -  2 z (1 + z) f_3 
    - (3d-8) z f_2 - (d-2) f_1 \,,
    \nonumber\\
  & z f'_6  - \frac{1}{2}(3d-10) f_6 = 0 \,,
    \nonumber\\
  & f'_5 = 0\,,
    \nonumber\\
  & z f'_4 - (d-3) f_4 = 0, 
    \nonumber\\
  & z f'_2 - z f_3 = 0\,,
    \nonumber\\
  &  z(z-1) f'_3 + \frac{1}{2}(d-3) (3d-8) f_2 - \frac{1}{2} (4 - d - 16 z + 5 d z) f_3 = 0\,,
    \nonumber\\
  & z f'_1 - \frac{1}{2}(d-2) f_1 = 0 \,,
    \label{eq:DNsys}
\end{align}
where $f'_j=df_j/dz$.

To solve the system in eq.~(\ref{eq:DNsys}), we substitute the following collective ansatz:
\begin{align}
  f_j = \sum_{n=0}^\infty
  \sum_{k=1}^{K} c_{j,n,k} z^{\mu_k+n} \,.
  \label{eq:ansatz}
\end{align}
The exponent shifts $\mu_k$ are determined as usual in the Frobenius method from the indicial polynomials.
Actually, it is easy to establish that $\mu_j$ can take the values $0,-\varepsilon,-2\varepsilon,-3\varepsilon$.
Therefore, we have, for each value of $j$, four different solutions $f_j^{(1)},f_j^{(2)},f_j^{(3)},f_j^{(4)}$,
corresponding to the different values of $\mu_k$, and the solution we are looking for is the linear
combination
\begin{align}
  f_j = \sum_{k=1}^4 C_{j,k} f_j^{(k)} \,,
  \label{eq:fj}
\end{align}
with unknown constants $C_{j,k}$, which should be determined from the boundary conditions at $z=0$.
Thus, for each value $j=1,\dots,7$, we need four boundary conditions, one for each value of $\mu_k$.

The boundary conditions correspond to the expansions of the integrals in eq.~(\ref{eq:DN_MI}) about $z=0$.
Following the standard rules of the large-mass expansion \cite{Smirnov:1990rz,Smirnov:1994tg},
we should take into account the four hard subgraphs $\{123456\}$, $\{23456\}$, $\{356\}+\{245\}$,
and $\{5\}$. These four subgraphs provide the four boundary conditions for
eq.~(\ref{eq:fj}).


\section{Results and discussion}
\label{sec:results}

We present the terms of the $\eps$ expansions analytically in terms of the bases $\Re H_j$
and $\Im H_j$, $j=1,\dots,6$ in the appendix. In each case, we take the prefactor in such
a way that the terms of the expansion are homogeneous in the weights. In some cases, this requires us to
evaluate additional integrals (with dots on lines) and to re-express the original integral
with the help of integration-by-parts relations. Moreover, we find that, with the suitable
choice of prefactors, the elements of the expansion are expressed in each case either through the
$\Re H$ basis or the $\Im H$ basis. This feature is a convenient property
which allows us to reduce the length of PSLQ vector. In addition to the analytic expressions,
we also give their numerical values accurate to 50 decimal figures. 

There is, of course, a certain degree of arbitrariness in the choice of the basis elements.
We just use the lexicographical ordering of the three-letter alphabet $\{-1,0,1\}$.
The sets of basis elements and the transformation between different bases (with arguments $\omega$
and $\omega^2$), as well as all analytic results can be found in the
attachment in a \texttt{Mathematica}-readable form. In addition, we give the numerical values of all basis
elements both for ${\cal A}_{H(\omega)}$ and ${\cal A}_{H(\omega^2)}$ to an accuracy of 20,000 decimal figures.

\subsection{Integrals evaluated in terms of $\Gamma$ functions}

Four of the integrals in figure~\ref{fig:tad1234} can be evaluated in terms of $\Gamma$ functions, namely
\begin{align}
  {\bf V_1} &= e^{\gamma\eps} \Gamma(-1+\eps) \,,
  \\
  {\bf T_{100}} &= e^{2\gamma\eps} \frac{\Gamma(-1+2\eps) \Gamma(\eps) \Gamma^2(1-\eps)}{\Gamma(2-\eps)} \,,
  \\
  {\bf E_1} &= e^{3\gamma\eps}\frac{\Gamma(2 - 3 \eps) \Gamma^4(1 - \eps) \Gamma^2(\eps) \Gamma(-1 + 3 \eps)}
              {\Gamma^2(2 - 2 \eps) \Gamma(2 - \eps)}  \,,
  \\
  {\bf BN_3} &= e^{3\gamma\eps}\frac{\Gamma^3(1-\eps) \Gamma(-1 + 2\eps) \Gamma(-2 + 3\eps)}{\Gamma(2 - \eps)} \,,
  \\
  {\bf BN_2} &= e^{3\gamma\eps}\frac{\Gamma^2(1 - \eps) \Gamma(\eps) \Gamma^2(-1 + 2 \eps) \Gamma(-2 + 3 \eps)}
               {\Gamma(2 - \eps) \Gamma(-2 + 4 \eps)} \,.
\end{align}

\subsection{Two-loop integral $\bf T_{111}$}

The two-loop integral $\bf T_{111}$ was considered in ref.~\cite{Davydychev:1992mt}, where its representation
in terms of the hypergeometric function ${}_4F_3$ was given. The construction of its
$\eps$ expansion was discussed in great detail in ref.~\cite{Davydychev:2000kw}. There, the expansion
to all orders in $\eps$ was found in terms of log-sine integrals. 

We find that $\bf T_{111}$ can be written in terms of our bases as
\begin{align}
  {\bf T_{111}} &=  \frac{e^{2\gamma\varepsilon}\, \Gamma^2(1+\varepsilon)}{\sqrt{3}(-1+\eps)(1-2\varepsilon)}
                  \left( \frac{3\sqrt{3}}{2\varepsilon^2} + \bar{T}_{111}^{(0)} + \varepsilon \bar{T}_{111}^{(1)}
                  + \varepsilon^2 \bar{T}_{111}^{(2)} + \varepsilon^3 \bar{T}_{111}^{(3)} + \varepsilon^4 \bar{T}_{111}^{(4)}
                  + \dots
                  \right) \,,
                  \label{eq:twoloop}
\end{align}
where $\bar{T}^{(k)}$ are expressed in terms of the homogeneous bases $\Im H_{k+2}$. 
They are presented in the appendix.

It should be noted here that, in eq.~(\ref{eq:twoloop}), it is necessary to take out the factor $\Gamma^2(1+\eps)$
to avoid the mixing of the $\Im H$ and $\Re H$ bases. This mixing occurs, since
$\zeta_k\in \Re{\cal A}_{H(\omega)}$, while $\zeta_k\notin \Im{\cal A}_{H(\omega)}$ 
and $\sqrt{3}\zeta_k\notin \Im{\cal A}_{H(\omega)}$.

To the accuracy of 50 decimal figures, we have
\begin{align}
  {\bf T_{111}} =  & - \frac{3}{2\eps^2} - \frac{9}{2\eps}
                     \nonumber\\
                   & -9.4515402422381513188062793166606221801853203373564
                     \nonumber\\
                   & -24.208928021203592678721338219570948925801493085960 \varepsilon
                     \nonumber\\
                   & -38.717599744915838872316613641777943709417494962957 \varepsilon^2
                     \nonumber\\
                   & -101.95399100959711442266247857697577872945662436761 \varepsilon^3 
                     \nonumber\\
                   & -152.48276547467415258599823439719064567392808286431 \varepsilon^4 + \dots\,.
                     \label{eq:J111num}
\end{align}

\subsection{Diagram $\bf BN$}

The integral $\bf BN$ belongs to the class of the so-called `QED-type' integrals. These are the
integrals with an even number of massive lines at each vertex. They have an especially simple structure
and were considered in ref.~\cite{Lee:2010hs} through weight six.
We have
\begin{align}
  {\bf BN} &=  \frac{e^{3\gamma\varepsilon}\, \Gamma^3(1+\varepsilon) }{(1-\eps)(1-2\eps)(1-3\eps)(2-3\eps)}
             \nonumber\\
           &\qquad\qquad
             \times  \left( \frac{4}{\eps^3} - \frac{44}{3\eps^2}
             + \eps \overline{BN}^{(1)} + \eps^2 \overline{BN}^{(2)} + \eps^3 \overline{BN}^{(3)} + \eps^4 \overline{BN}^{(4)} 
             + \dots
             \right) \,,
             \label{eq:BNint}
\end{align}
where $\overline{BN}^{(k)}$ are expressed in terms of the homogeneous bases $\Re H_{k+2}$ and
are explicitly given in the appendix.

Numerically, we have
\begin{align}
  {\bf BN} =  &  \frac{2}{\eps^3} + \frac{23}{3\eps^2}
                \nonumber\\
              & +22.434802200544679309417245499938075567656849703620 \frac{1}{\eps}
                \nonumber\\
              & +39.429294629102082115299964760073056361154617945864
                \nonumber\\
              & +62.927093755359100705477989920486998916962128778534 \eps
                \nonumber\\
              & -126.33666539901207007224982170333707283310295118164 \eps^2
                \nonumber\\
              & -584.77850194492638360751899973098599972365977374517 \eps^3
                \nonumber\\
              & -4108.8159602165199632668134484734533423368833382544 \eps^4 + \dots\,.
                \label{eq:B4num}
\end{align}

\subsection{Diagram $\bf BN_1$}

The diagram $\bf BN_1$ was previously considered in refs.~\cite{Fleischer:1999mp,Davydychev:2000na}. There,
its explicit representation in terms of the hypergeometric function ${}_QF_P$ of argument $1/4$
was obtained. The corresponding $\eps$ expansion, through weight-five polylogarithms, 
was constructed in terms of log-sine integrals.

In order to keep the property of the weight homogeneity and to separate the
real and imaginary bases, we write $\bf BN_1$ in terms of the
additional integrals $\bf BN_1^\prime$, which is $\bf BN_1$ with
additional dots and $\bf V_1$,
\begin{align}
  {\bf BN_1} = &   \frac{9 \left( \overline{BN}_1^{(0)}
                 + \eps \overline{BN}_1^{(1)} + \eps^2 \overline{BN}_1^{(2)} + \eps^3 \overline{BN}_1^{(3)} + \eps^4 \overline{BN}_1^{(4)} 
                 + \dots
                 \right)}{2\sqrt{3}(1-\eps)(-1 + 2\eps)(-2 + 3\eps)(-1 + 3\eps)} 
                 \nonumber\\
               & + \frac{(-1 + \eps)^2(-4 + 15\eps)}{2(-1 + 2\eps)(-2 + 3\eps)(-1 + 3\eps)} {\bf V_1}^3 \,,
\end{align}

Numerically, we have
\begin{align}
  {\bf BN_1} &=  \frac{1}{\eps^3} + \frac{15}{4\eps^2} 
               + \frac{1}{\eps}\left( \frac{65}{8} + \frac{3}{2}\zeta_2 \right)
               \nonumber\\
             &+21.761988509912961923611112300835065104056263673628
               \nonumber\\
             &+7.8517428364255311757755955915256850308098967304537 \eps
               \nonumber\\
             &-71.052070175912095038002576965797146234900221815177 \eps^2
               \nonumber\\
             &-716.82162754590202527205013899706978778657937882568 \eps^3
               \nonumber\\
             &-2486.5823094068232493409812154539302262366876271196 \eps^4 + \dots\,.
\end{align}


\subsection{Diagram $\bf E_3$}

The integral $\bf E_3$ was also considered in refs.~\cite{Fleischer:1999mp,Davydychev:2000na}, where the same
analysis as for $\bf BN_1$ can be found. Our representation of this integral reads:
\begin{align}
  {\bf E_3} &= \frac{1}{\sqrt{3}(1-\eps)(1-2\eps)^2} \Big(
              \frac{1}{\eps} \bar{E}_3^{(-1)}
              + \bar{E}_3^{(0)} + \eps \bar{E}_3^{(1)} + \eps^2 \bar{E}_3^{(2)} + \eps^3 \bar{E}_3^{(3)} + \dots \Big)
              \nonumber\\ 
            & + \frac{e^{3\gamma\eps} \Gamma(1-\eps) \Gamma(1+2\eps) \Gamma^2(1+\eps)}{2\eps^3 (1-\eps)(1-2\eps)^2}
              \left(
              1 + \frac{1-\eps}{3(1-3\eps)}\frac{\Gamma(1+2\eps) \Gamma(1+3\eps)}{\Gamma(1+\eps) \Gamma(1+4\eps)}
              \right) \,,
\end{align}
where $\bar{E}_3^{(k)}$ are expressed in terms of the homogeneous bases $\Im H_{k+3}$ and
are explicitly given in the appendix.

Numerically, we have
\begin{align}
  {\bf E_3} &= \frac{2}{3\eps^3} + \frac{11}{3\eps^2}
              \nonumber\\
            &+13.774007275662264537042486899983634774794795287960 \frac{1}{\eps}
              \nonumber\\
            &+55.659622461206330171361395086424121982538065496717
              \nonumber\\
            &+151.93523620176745531459840301896170801637068885020 \eps
              \nonumber\\
            &+574.65405761296725286615112197885868312967183890132 \eps^2
              \nonumber\\
              &+1417.6830429498080864815903343175553853260055772458 \eps^3 +
              \dots\,.
\end{align}

\subsection{$\bf D$-type diagrams}

For all diagrams of the mercedes type, we have the same representation
\begin{align}
  {\bf D}_x = \frac{1}{(1-\eps)(1-2\eps)}\left( \frac{2\zeta_3}{\eps}
  + \bar{D}_x^{(0)} + \eps \bar{D}_x^{(1)} + \eps^2 \bar{D}_x^{(2)} + \dots \right) \,,
\end{align}
where all $\bar{D}_x^{(k)}$ are expressed in terms of the homogeneous bases $\Re H_{k+4}$ and
are explicitly given in the appendix.

It should be noted that the integral $\bf D_5$ is sometimes replaced by 
the fully massive three-loop integral with 5 lines. The latter integral was considered in
ref.~\cite{Kalmykov:2005hb},
where, again, the hypergeometric representation was presented.

Numerically, we have
\begin{align}
  {\bf D_6} &=  \frac{2\zeta_3}{\eps} 
              \nonumber\\
            &-10.035278479768789171914700685158900238650333496003
              \nonumber\\
            &+41.876702083031576174334902670970991466431593917007 \eps
              \nonumber\\
            &-146.80128953959941603962375965123680914875429640084 \eps^2 + \dots \,,
              \label{eq:D6num}
  \\
  {\bf D_5} &=  \frac{2\zeta_3}{\eps} 
              \nonumber\\
            &-8.2168598175087380629133983386010858249695083391726
              \nonumber\\
            &+36.473684211550968259944718569763658485586503339935 \eps
              \nonumber\\
            &-122.50284392807361438626452778528813541284319434062 \eps^2 + \dots \,,
              \label{eq:D5num}
  \\
  {\bf D_4} &=  \frac{2\zeta_3}{\eps} 
              \nonumber\\
            &-5.9132047838840205304957178925354050268834109915340
              \nonumber\\
            &+31.793875865203350915027031305982932318904242706405 \eps
              \nonumber\\
            &-95.531868585060481541530996460991495963507326204506 \eps^2 + \dots \,,
              \label{eq:D4num}
  \\
  {\bf D_3} &=  \frac{2\zeta_3}{\eps} 
              \nonumber\\
            &-3.0270094939876520197863747017589572861507417864174
              \nonumber\\
            &+28.736435119523636809010469958622996315186640636723 \eps
              \nonumber\\
            &-63.461003588316921857688768719288636938911285487969 \eps^2 + \dots \,,
              \label{eq:D3num}
  \\
  {\bf D_M} &=  \frac{2\zeta_3}{\eps} 
              \nonumber\\
            &-2.8608622241393273502727845677732419175614414620201
              \nonumber\\
            &+29.006674437837759083319026315817224175180046773850 \eps
              \nonumber\\
            &-62.361396342296251606481070393459830940783578107024 \eps^2 + \dots \,,
              \label{eq:DMnum}
  \\
  {\bf D_N} &=  \frac{2\zeta_3}{\eps} 
              \nonumber\\
            &+1.1202483970392420822725165482242095262757766719791
              \nonumber\\
            &+0.681035275345890550785882982356275565304510792838 \eps
              \nonumber\\
            &-13.303460640858248361888942340988906552580290997554 \eps^2 + \dots \,.
              \label{eq:DNnum}
\end{align}


\section{Conclusions}

In this work, we considered the three-loop massive vacuum bubble diagrams and
constructed the pertaining bases of irrational constants through weight six,
which are harmonic polylogarithms of argument $\omega=e^{i\pi/3}$.
These bases are smaller than the bases
of the algebra of the sixth root of unity. Nevertheless, we found by explicit calculation that
such reduced bases are large enough to describe all the three-loop single-scale vacuum integrals.
We presented the results for all relevant master integrals both numerically and analytically
in terms of the introduced constants.
Our basis is universal, and its application is not restricted to
three-loop tadpoles. As an example, we succeeded in
reconstructing\footnote{Our results are available online as an attachment
  to the arXiv version of this paper.} all the
three-loop integrals contributing to the massive planar form factor through weight six
presented in ref.~\cite{Henn:2016kjz}. As expected, these integrals only
involve the real parts of the basis $\Re {\cal A}_{H(\omega)}$.

\acknowledgments

We thank  Roman Lee for his help with the \texttt{SummerTime} package.
This work was supported in part by the German Federal Ministry for Education
and Research BMBF through Grant No.\ 05H15GUCC1, by the German Research
Foundation DFG through the Collaborative Research Centre No.\ SFB~676
{\it Particles, Strings and the Early Universe: the Structure of Matter and                                                                 
  Space-Time},
and by the Heisenberg--Landau Programme.


\appendix

\section{\boldmath Master integrals in terms of harmonic
  polylogarithms of argument $e^{\frac{i \pi}{3}}$}
\label{sec:MIsToHPL}

In this appendix, we present the first few terms of the $\eps$ expansions of the relevant master integrals.

For the elements of the bases $\Re H_k$ and $\Im H_k$, we introduce the following short-hand notation:
\begin{align}
  \Re H_{n_1\dots n_w}(\omega) &= {\cal R}_{n_1\dots n_w} \,,
  \\
  \Im H_{n_1\dots n_w}(\omega) &= {\cal I}_{n_1\dots n_w} \,.
\end{align}
Then, the coefficients $\bar{T}$, $\bar{B}$, $\bar{E}$, and $\bar{D}$
introduced in section~\ref{sec:results} read:
\begin{align}
  \bar{D}_6^{(0)} = &  -\frac{72}{11} \HR_{1,-1,1,0}+\frac{180}{11}
                      \HR_{1,-1,1,1}+\frac{148}{11}
                      \HR_{1,0,1,0}-\frac{144}{11}
                      \HR_{1,1,-1,0}+\frac{360}{11}
                      \HR_{1,1,-1,1}\nonumber\\
                    & +\frac{540}{11} \HR_{1,1,1,-1}-\frac{33587}{55} \HR_{1,1,1,1}\\
  \bar{D}_6^{(1)} = & 156 \HR_{1,-1,1,1,0}-16 \HR_{1,0,-1,1,0}-16
                      \HR_{1,0,1,-1,0}-468
                      \HR_{1,0,1,1,1}+\frac{7712}{87}
                      \HR_{1,1,-1,0,0}\nonumber\\
                    & -\frac{7084}{87} \HR_{1,1,-1,0,1}+\frac{15884}{87}
                      \HR_{1,1,-1,1,0}+\frac{5632}{29}
                      \HR_{1,1,-1,1,1}-36
                      \HR_{1,1,0,1,-1}\nonumber\\
                    & -\frac{13708319
                      \HR_{1,1,0,1,1}}{15138}+ \frac{9448}{87} \HR_{1,1,1,-1,0}+\frac{15640}{29}
                      \HR_{1,1,1,-1,1}-108
                      \HR_{1,1,1,0,-1}\nonumber\\
                    & - \frac{59151355 \HR_{1,1,1,0,1}}{45414}+992
                      \HR_{1,1,1,1,-1} - \frac{51713387 \HR_{1,1,1,1,0}}{22707}\\
  \bar{D}_6^{(2)} = &  -624 \HR_{1,-1,-1,1,1,0}-\frac{8592}{11}
                      \HR_{1,-1,1,-1,1,0}+\frac{4320}{11}
                      \HR_{1,-1,1,-1,1,1}+\frac{2848}{11}
                      \HR_{1,-1,1,0,1,0}\nonumber\\
                    & -\frac{10320}{11}
                      \HR_{1,-1,1,1,-1,0}+\frac{8640}{11} \HR_{1,-1,1,1,-1,1}+144
                      \HR_{1,-1,1,1,0,-1}+\frac{12960}{11}
                      \HR_{1,-1,1,1,1,-1}\nonumber\\
                    & -\frac{391256}{55} \HR_{1,-1,1,1,1,1}+32
                      \HR_{1,0,-1,-1,1,0}+32 \HR_{1,0,-1,1,-1,0}+1976
                      \HR_{1,0,-1,1,1,1}\nonumber\\
                    & +32 \HR_{1,0,1,-1,-1,1}+32 \HR_{1,0,1,-1,1,-1}+\frac{5024}{33} \HR_{1,0,1,-1,1,0}+\frac{17432}{11}
                      \HR_{1,0,1,-1,1,1}\nonumber\\
                    & -\frac{53444}{99} \HR_{1,0,1,0,1,1}+64 \HR_{1,0,1,1,-1,-1}+\frac{11104}{33} \HR_{1,0,1,1,-1,0}+\frac{12776}{11} \HR_{1,0,1,1,-1,1}\nonumber\\
                    & -\frac{39451}{99}
                      \HR_{1,0,1,1,0,1}+\frac{7768}{11} \HR_{1,0,1,1,1,-1}+\frac{96931}{99} \HR_{1,0,1,1,1,0}-\frac{16128}{11} \HR_{1,1,-1,-1,1,0}\nonumber\\
                    & +\frac{6000}{11}
                      \HR_{1,1,-1,-1,1,1}-\frac{19584}{11} \HR_{1,1,-1,1,-1,0}+\frac{14640}{11} \HR_{1,1,-1,1,-1,1}+288 \HR_{1,1,-1,1,0,-1}\nonumber\\
                    & -\frac{5696}{11} \HR_{1,1,-1,1,0,1}+\frac{23280}{11}
                      \HR_{1,1,-1,1,1,-1}-\frac{2128}{11} \HR_{1,1,-1,1,1,0}-\frac{372192}{55} \HR_{1,1,-1,1,1,1}\nonumber\\
                    & +\frac{12064}{3} \HR_{1,1,0,-1,1,1}+208 \HR_{1,1,0,1,-1,-1}+\frac{1405771580504
                      \HR_{1,1,0,1,-1,0}}{52401195}\nonumber\\
                    & -\frac{449711597968
                      \HR_{1,1,0,1,-1,1}}{17467065} -\frac{820125684944 \HR_{1,1,0,1,1,-1}}{17467065}-\frac{44309295476
                      \HR_{1,1,1,-1,-1,0}}{1940785}\nonumber\\
                    & +\frac{42581643956 \HR_{1,1,1,-1,-1,1}}{1940785}-\frac{5328}{5} \HR_{1,1,1,-1,0,-1}+\frac{56916979364
                      \HR_{1,1,1,-1,0,0}}{7485885}\nonumber\\
                    & -\frac{113890854256 \HR_{1,1,1,-1,0,1}}{7485885}+\frac{44868241556 \HR_{1,1,1,-1,1,-1}}{1940785}-\frac{42534744232
                      \HR_{1,1,1,-1,1,0}}{52401195}\nonumber\\
                    & +\frac{104647550068 \HR_{1,1,1,-1,1,1}}{17467065}+432 \HR_{1,1,1,0,-1,-1}+\frac{179716135264 \HR_{1,1,1,0,-1,0}}{1940785}\nonumber\\
                    & -\frac{26026092432
                      \HR_{1,1,1,0,-1,1}}{277255}+\frac{37024}{5} \HR_{1,1,1,0,0,-1}+\frac{2350091860163 \HR_{1,1,1,0,0,1}}{471610755}\nonumber\\
                    & -\frac{909009749728
                      \HR_{1,1,1,0,1,-1}}{5822355}-\frac{1515749729887 \HR_{1,1,1,0,1,0}}{471610755}+\frac{163974915824 \HR_{1,1,1,1,-1,-1}}{1940785}\nonumber\\
                    & +\frac{551982130556
                      \HR_{1,1,1,1,-1,0}}{17467065}+\frac{280123847084 \HR_{1,1,1,1,-1,1}}{17467065}-\frac{19847388288 \HR_{1,1,1,1,0,-1}}{55451}\nonumber\\
                    & +\frac{8283679238293
                      \HR_{1,1,1,1,0,0}}{419209560}-\frac{4590966964
                      \HR_{1,1,1,1,1,-1}}{388157} - \frac{14507424807748399
                      \HR_{1,1,1,1,1,1}}{2400927480}
\end{align}

\begin{align}
  \bar{D}_5^{(0)} = & -26 \HR_{-1,1,1,0}-26 \HR_{1,-1,1,0}
                      +\frac{16}{3} \HR_{1,0,1,0} 
                      -26 \HR_{1,1,-1,0} +6 \HR_{1,1,0,-1} \nonumber\\
                    & +\frac{616}{15} \HR_{1,1,1,1}\\
  \bar{D}_5^{(1)} = & 104 \HR_{-1,-1,1,1,0}+\frac{1432}{11}
                      \HR_{-1,1,-1,1,0}-\frac{720}{11}
                      \HR_{-1,1,-1,1,1} -\frac{1424}{33} \HR_{-1,1,0,1,0}\nonumber\\
                    & +\frac{1720}{11} \HR_{-1,1,1,-1,0}-\frac{1440}{11}
                      \HR_{-1,1,1,-1,1}-24 \HR_{-1,1,1,0,-1} -\frac{2160}{11} \HR_{-1,1,1,1,-1}\nonumber\\
                    & +\frac{195628}{165} \HR_{-1,1,1,1,1}+260 \HR_{1,-1,-1,1,1}+\frac{2572}{11}
                      \HR_{1,-1,1,-1,1} +\frac{668}{33} \HR_{1,-1,1,0,1}\nonumber\\
                    & +\frac{2284}{11}
                      \HR_{1,-1,1,1,-1}+\frac{674}{3}
                      \HR_{1,-1,1,1,0}+\frac{1076}{11}
                      \HR_{1,0,-1,1,0} +\frac{1316}{11}
                      \HR_{1,0,1,-1,0}\nonumber\\
                    & -\frac{360}{11} \HR_{1,0,1,-1,1}-20
                      \HR_{1,0,1,0,-1}-\frac{720}{11}
                      \HR_{1,0,1,1,-1} -\frac{14198}{33} \HR_{1,0,1,1,1}\nonumber\\
                    & +\frac{4864}{11}
                      \HR_{1,1,-1,-1,1}-\frac{11992}{405}
                      \HR_{1,1,-1,0,0}+\frac{759664
                      \HR_{1,1,-1,0,1}}{4455} +\frac{5008}{11} \HR_{1,1,-1,1,-1}\nonumber\\
                    & +\frac{2357524
                      \HR_{1,1,-1,1,0}}{4455}-\frac{198584}{135}
                      \HR_{1,1,-1,1,1} -24 \HR_{1,1,0,-1,-1} -\frac{926}{11} \HR_{1,1,0,1,-1}\nonumber\\
                    & -\frac{559454947 \HR_{1,1,0,1,1}}{775170} +\frac{6012}{11}
                      \HR_{1,1,1,-1,-1}+\frac{850486}{891}
                      \HR_{1,1,1,-1,0} -\frac{5116084 \HR_{1,1,1,-1,1}}{1485}\nonumber\\
                    & +\frac{3514}{33} \HR_{1,1,1,0,-1}-\frac{44410145
                      \HR_{1,1,1,0,1}}{42282}-\frac{9118444
                      \HR_{1,1,1,1,-1}}{1485}-\frac{2126716774
                      \HR_{1,1,1,1,0}}{1162755}\\
  \bar{D}_5^{(2)} = & -416 \HR_{-1,-1,-1,1,1,0}-\frac{5728}{11}
                      \HR_{-1,-1,1,-1,1,0}+\frac{2880}{11}
                      \HR_{-1,-1,1,-1,1,1} +\frac{5696}{33} \HR_{-1,-1,1,0,1,0}\nonumber\\
                    & -\frac{6880}{11}
                      \HR_{-1,-1,1,1,-1,0}+\frac{5760}{11}\HR_{-1,-1,1,1,-1,1}
                      + 96 \HR_{-1,-1,1,1,0,-1} +\frac{8640}{11} \HR_{-1,-1,1,1,1,-1}\nonumber\\
                    & -\frac{782512}{165} \HR_{-1,-1,1,1,1,1}-1040
                      \HR_{-1,1,-1,-1,1,1} -\frac{10288}{11}
                      \HR_{-1,1,-1,1,-1,1} -\frac{2672}{33} \HR_{-1,1,-1,1,0,1}\nonumber\\
                    & -\frac{9136}{11} \HR_{-1,1,-1,1,1,-1}-\frac{2696}{3}
                      \HR_{-1,1,-1,1,1,0}-\frac{4304}{11}
                      \HR_{-1,1,0,-1,1,0} -\frac{5264}{11} \HR_{-1,1,0,1,-1,0}\nonumber\\
                    & +\frac{1440}{11} \HR_{-1,1,0,1,-1,1}+80 \HR_{-1,1,0,1,0,-1}+\frac{2880}{11}
                      \HR_{-1,1,0,1,1,-1}+\frac{56792}{33} \HR_{-1,1,0,1,1,1}\nonumber\\
                    & -\frac{19456}{11}
                      \HR_{-1,1,1,-1,-1,1}+\frac{3257312
                      \HR_{-1,1,1,-1,0,0}}{11745} -\frac{117630944
                      \HR_{-1,1,1,-1,0,1}}{129195}\nonumber\\
                    & -\frac{20032}{11}
                      \HR_{-1,1,1,-1,1,-1}-\frac{302982704
                      \HR_{-1,1,1,-1,1,0}}{129195}+\frac{24202144
                      \HR_{-1,1,1,-1,1,1}}{3915}\nonumber\\
                    & +96
                      \HR_{-1,1,1,0,-1,-1}+\frac{3704}{11}
                      \HR_{-1,1,1,0,1,-1} +\frac{38639896921 \HR_{-1,1,1,0,1,1}}{11239965}\nonumber\\
                    & -\frac{24048}{11} \HR_{-1,1,1,1,-1,-1} -\frac{123547352
                      \HR_{-1,1,1,1,-1,0}}{25839} +\frac{649919504 \HR_{-1,1,1,1,-1,1}}{43065}\nonumber\\
                    & -\frac{14056}{33}
                      \HR_{-1,1,1,1,0,-1}+\frac{3586576799
                      \HR_{-1,1,1,1,0,1}}{613089} +\frac{41605936
                      \HR_{-1,1,1,1,1,-1}}{1485}\nonumber\\
                    & +\frac{357300160814
                      \HR_{-1,1,1,1,1,0}}{33719895}-1040
                      \HR_{1,-1,-1,-1,1,1} -\frac{10288}{11} \HR_{1,-1,-1,1,-1,1}\nonumber\\
                    & -\frac{2672}{33}
                      \HR_{1,-1,-1,1,0,1}-\frac{9136}{11}
                      \HR_{1,-1,-1,1,1,-1} -\frac{2696}{3} \HR_{1,-1,-1,1,1,0}\nonumber\\
                    & -\frac{576}{11} \HR_{1,-1,1,-1,1,-1}-\frac{387824}{121}
                      \HR_{1,-1,1,-1,1,0} +\frac{13241792 \HR_{1,-1,1,-1,1,1}}{1815}\nonumber\\
                    & -\frac{35584}{33}
                      \HR_{1,-1,1,0,-1,1}-\frac{28792}{33}
                      \HR_{1,-1,1,0,1,-1} +\frac{2296288
                      \HR_{1,-1,1,0,1,0}}{1089}\nonumber\\
                    & +\frac{14864}{11}
                      \HR_{1,-1,1,1,-1,-1}-\frac{5392193704
                      \HR_{1,-1,1,1,-1,0}}{1421145} +\frac{5442900016 \HR_{1,-1,1,1,-1,1}}{473715}\nonumber\\
                    & -\frac{32280}{11}
                      \HR_{1,-1,1,1,0,-1}+\frac{6570705136
                      \HR_{1,-1,1,1,1,-1}}{473715} +\frac{25996220167837 \HR_{1,-1,1,1,1,1}}{1854594225}\nonumber\\
                    & +\frac{64}{3} \HR_{1,0,-1,-1,1,0}+\frac{64}{3}
                      \HR_{1,0,-1,1,-1,0}-\frac{208}{3}
                      \HR_{1,0,-1,1,1,1} -\frac{47936}{33} \HR_{1,0,1,-1,-1,1}\nonumber\\
                    & -\frac{47936}{33} \HR_{1,0,1,-1,1,-1}+\frac{540736}{363}
                      \HR_{1,0,1,-1,1,0}-\frac{345216128 \HR_{1,0,1,-1,1,1}}{94743}\nonumber\\
                    & +80 \HR_{1,0,1,0,-1,-1}+\frac{49280509019
                      \HR_{1,0,1,0,1,1}}{148367538} -\frac{95872}{33}
                      \HR_{1,0,1,1,-1,-1}\nonumber\\
                    & +\frac{634929920
                      \HR_{1,0,1,1,-1,0}}{852687}-\frac{676898224
                      \HR_{1,0,1,1,-1,1}}{284229} +\frac{387546944255 \HR_{1,0,1,1,0,1}}{222551307}\nonumber\\
                    & +\frac{352181312
                      \HR_{1,0,1,1,1,-1}}{94743}+\frac{4241416264027
                      \HR_{1,0,1,1,1,0}}{445102614} +\frac{19456}{11} \HR_{1,1,-1,-1,-1,1}\nonumber\\
                    & +\frac{18880}{11} \HR_{1,1,-1,-1,1,-1}-\frac{1901200}{363}
                      \HR_{1,1,-1,-1,1,0}+\frac{19253152 \HR_{1,1,-1,-1,1,1}}{1815}\nonumber\\
                    & +\frac{824}{11} \HR_{1,1,-1,0,1,-1}+\frac{53776}{11} \HR_{1,1,-1,1,-1,-1}-\frac{5754526792
                      \HR_{1,1,-1,1,-1,0}}{1421145}\nonumber\\
                    & +\frac{965522608
                      \HR_{1,1,-1,1,-1,1}}{94743}-\frac{34744}{11}
                      \HR_{1,1,-1,1,0,-1} -\frac{1791826416874
                      \HR_{1,1,-1,1,0,1}}{370918845}\nonumber\\
                    & +\frac{201931024 \HR_{1,1,-1,1,1,-1}}{94743}
                      -\frac{5946278372713
                      \HR_{1,1,-1,1,1,0}}{370918845} +\frac{181225331611139\HR_{1,1,-1,1,1,1}}{1854594225}\nonumber\\
                    & +96 \HR_{1,1,0,-1,-1,-1}+\frac{824}{11}
                      \HR_{1,1,0,-1,1,-1} - \frac{61094644559
                      \HR_{1,1,0,-1,1,1}}{11239965} -\frac{93400}{33}
                      \HR_{1,1,0,1,-1,-1}\nonumber\\
                    & +\frac{71673534978119191589 \HR_{1,1,0,1,-1,0}}{588987597758175}-\frac{27778395373685590393 \HR_{1,1,0,1,-1,1}}{196329199252725}\nonumber\\
                    & -\frac{39808709428404512639
                      \HR_{1,1,0,1,1,-1}}{196329199252725}+\frac{92688}{11} \HR_{1,1,1,-1,-1,-1}\nonumber\\
                    & -\frac{49767882477121799173 \HR_{1,1,1,-1,-1,0}}{392658398505450}+\frac{50254587600763374613
                      \HR_{1,1,1,-1,-1,1}}{392658398505450}\nonumber\\
                    & -\frac{191463192286 \HR_{1,1,1,-1,0,-1}}{18733275}+\frac{30945325345674631 \HR_{1,1,1,-1,0,0}}{692519221350}\nonumber\\
                    & -\frac{15249629583930690367
                      \HR_{1,1,1,-1,0,1}}{168282170788050} +\frac{43031534692552744213 \HR_{1,1,1,-1,1,-1}}{392658398505450}\nonumber\\
                    & -\frac{41886306721880707259
                      \HR_{1,1,1,-1,1,0}}{1177975195516350}+\frac{108791082954484641911 \HR_{1,1,1,-1,1,1}}{392658398505450}\nonumber\\
                    & -\frac{14056}{33} \HR_{1,1,1,0,-1,-1}+\frac{89657794477349654246
                      \HR_{1,1,1,0,-1,0}}{196329199252725}\nonumber\\
                    & -\frac{4767577759273275116 \HR_{1,1,1,0,-1,1}}{9349009488225}+\frac{1559414393279 \HR_{1,1,1,0,0,-1}}{33719895}\nonumber\\
                    & +\frac{13690397558977227041
                      \HR_{1,1,1,0,0,1}}{848142140771772}-\frac{15533573816964260686 \HR_{1,1,1,0,1,-1}}{21814355472525}\nonumber\\
                    & -\frac{882309482505750318589
                      \HR_{1,1,1,0,1,0}}{21203553519294300}+\frac{88944543800158224026 \HR_{1,1,1,1,-1,-1}}{196329199252725}\nonumber\\
                    & +\frac{54362172151172329627
                      \HR_{1,1,1,1,-1,0}}{392658398505450}+\frac{187944687444042632053 \HR_{1,1,1,1,-1,1}}{392658398505450}\nonumber\\
                    & -\frac{557811878457413554
                      \HR_{1,1,1,1,0,-1}}{322379637525}-\frac{355643681672062658107 \HR_{1,1,1,1,0,0}}{18847603128261600}\nonumber\\
                    & +\frac{13865776375815235211
                      \HR_{1,1,1,1,1,-1}}{26177226567030}-\frac{3656464529968891203697567 \HR_{1,1,1,1,1,1}}{107945363370952800}
\end{align}

\begin{align}
  \bar{D}_4^{(0)} = & 12 \HR_{1,0,1,0}-\frac{2742}{5} \HR_{1,1,1,1}\\
  \bar{D}_4^{(1)} = &  -56 \HR_{1,0,-1,1,0}-56 \HR_{1,0,1,-1,0}+198
                      \HR_{1,0,1,1,1} +\frac{15539}{174} \HR_{1,1,0,1,1}\nonumber\\
                    & -\frac{152353}{522} \HR_{1,1,1,0,1}-\frac{219164}{261} \HR_{1,1,1,1,0}\\
  \bar{D}_4^{(2)} = & 304 \HR_{1,0,-1,-1,1,0}+304
                      \HR_{1,0,-1,1,-1,0}-1020 \HR_{1,0,-1,1,1,1} +304 \HR_{1,0,1,-1,-1,1}\nonumber\\
                    & +304 \HR_{1,0,1,-1,1,-1}+\frac{608}{3} \HR_{1,0,1,-1,1,0}-1628
                      \HR_{1,0,1,-1,1,1}+\frac{2602}{9} \HR_{1,0,1,0,1,1}\nonumber\\
                    & +608 \HR_{1,0,1,1,-1,-1}+\frac{2128}{3} \HR_{1,0,1,1,-1,0}-2540 \HR_{1,0,1,1,-1,1}+\frac{3733}{18}
                      \HR_{1,0,1,1,0,1}\nonumber\\
                    & -3756 \HR_{1,0,1,1,1,-1}-\frac{18773}{18} \HR_{1,0,1,1,1,0}-\frac{4144}{3} \HR_{1,1,0,-1,1,1}+608 \HR_{1,1,0,1,-1,-1}\nonumber\\
                    & -\frac{84256336
                      \HR_{1,1,0,1,-1,0}}{136107}-\frac{69730720 \HR_{1,1,0,1,-1,1}}{45369}-\frac{104376080 \HR_{1,1,0,1,1,-1}}{45369}\nonumber\\
                    & +\frac{17755424 \HR_{1,1,1,-1,-1,0}}{45369}-\frac{17755424
                      \HR_{1,1,1,-1,-1,1}}{45369}-\frac{21428960 \HR_{1,1,1,-1,0,0}}{136107}\nonumber\\
                    & +\frac{4285792 \HR_{1,1,1,-1,0,1}}{15123}-\frac{17755424 \HR_{1,1,1,-1,1,-1}}{45369}+\frac{3061280
                      \HR_{1,1,1,-1,1,0}}{136107}\nonumber\\
                    & -\frac{6734816 \HR_{1,1,1,-1,1,1}}{45369}-\frac{20198672 \HR_{1,1,1,0,-1,0}}{5041}+\frac{129077548 \HR_{1,1,1,0,-1,1}}{45369}\nonumber\\
                    & -\frac{1187408470
                      \HR_{1,1,1,0,0,1}}{1224963}+\frac{190647580 \HR_{1,1,1,0,1,-1}}{45369}+\frac{3293614283 \HR_{1,1,1,0,1,0}}{1224963}\nonumber\\
                    & -\frac{71021696 \HR_{1,1,1,1,-1,-1}}{45369}-\frac{26327008
                      \HR_{1,1,1,1,-1,0}}{45369}-\frac{7959328 \HR_{1,1,1,1,-1,1}}{45369}\nonumber\\
                    & +\frac{707222944 \HR_{1,1,1,1,0,-1}}{45369}+\frac{4181188831 \HR_{1,1,1,1,0,0}}{1088856}+\frac{3061280
                      \HR_{1,1,1,1,1,-1}}{5041}\nonumber\\
                    & +\frac{25498178108029 \HR_{1,1,1,1,1,1}}{68597928}
\end{align}

\begin{align}
  \bar{D}_3^{(0)} = & 12 \HR_{1,0,1,0}-\frac{2454}{5} \HR_{1,1,1,1}\\
  \bar{D}_3^{(1)} = & -72 \HR_{1,0,-1,1,0}-72 \HR_{1,0,1,-1,0}+234 \HR_{1,0,1,1,1}+\frac{8565}{58} \HR_{1,1,0,1,1}\nonumber\\
                    & -\frac{12469}{58} \HR_{1,1,1,0,1}-\frac{21182}{29} \HR_{1,1,1,1,0}\\
  \bar{D}_3^{(2)} = &  432 \HR_{1,0,-1,-1,1,0}+432 \HR_{1,0,-1,1,-1,0}-1404 \HR_{1,0,-1,1,1,1}+432 \HR_{1,0,1,-1,-1,1}\nonumber\\
                    & +432 \HR_{1,0,1,-1,1,-1}+288 \HR_{1,0,1,-1,1,0}-2268
                      \HR_{1,0,1,-1,1,1}+426 \HR_{1,0,1,0,1,1}\nonumber\\
                    & +864 \HR_{1,0,1,1,-1,-1}+1008 \HR_{1,0,1,1,-1,0}-3564 \HR_{1,0,1,1,-1,1}+\frac{597}{2} \HR_{1,0,1,1,0,1}\nonumber\\
                    & -5292
                      \HR_{1,0,1,1,1,-1}-\frac{2997}{2} \HR_{1,0,1,1,1,0}-1872 \HR_{1,1,0,-1,1,1}+864 \HR_{1,1,0,1,-1,-1}\nonumber\\
                    & -\frac{4434544 \HR_{1,1,0,1,-1,0}}{5041}-\frac{10551648
                      \HR_{1,1,0,1,-1,1}}{5041}-\frac{16021968 \HR_{1,1,0,1,1,-1}}{5041}\nonumber\\
                    & +\frac{2803488 \HR_{1,1,1,-1,-1,0}}{5041}-\frac{2803488 \HR_{1,1,1,-1,-1,1}}{5041}-\frac{1127840
                      \HR_{1,1,1,-1,0,0}}{5041}\nonumber\\
                    & +\frac{2030112 \HR_{1,1,1,-1,0,1}}{5041}-\frac{2803488 \HR_{1,1,1,-1,1,-1}}{5041}+\frac{161120 \HR_{1,1,1,-1,1,0}}{5041}\nonumber\\
                    & -\frac{1063392
                      \HR_{1,1,1,-1,1,1}}{5041}-\frac{28703376 \HR_{1,1,1,0,-1,0}}{5041}+\frac{21068364 \HR_{1,1,1,0,-1,1}}{5041}\nonumber\\
                    & -\frac{64500970 \HR_{1,1,1,0,0,1}}{45369}+\frac{30789948
                      \HR_{1,1,1,0,1,-1}}{5041}+\frac{171036689 \HR_{1,1,1,0,1,0}}{45369}\nonumber\\
                    & -\frac{11213952 \HR_{1,1,1,1,-1,-1}}{5041}-\frac{4156896 \HR_{1,1,1,1,-1,0}}{5041}-\frac{1256736
                      \HR_{1,1,1,1,-1,1}}{5041}\nonumber\\
                    & +\frac{112583712 \HR_{1,1,1,1,0,-1}}{5041}+\frac{216348149 \HR_{1,1,1,1,0,0}}{40328}+\frac{4350240 \HR_{1,1,1,1,1,-1}}{5041}\nonumber\\
                    & +\frac{1158178961791
                      \HR_{1,1,1,1,1,1}}{2540664}
\end{align}

\begin{align}
  \bar{D}_M^{(0)} = & 8 \HR_{1,0,1,0}-392 \HR_{1,1,1,1}\\
  \bar{D}_M^{(1)} = & -16 \HR_{1,0,-1,1,0}-16 \HR_{1,0,1,-1,0}+4
                      \HR_{1,0,1,1,1} -\frac{21209}{174}
                      \HR_{1,1,0,1,1}\nonumber\\
                    & -\frac{192041}{522}
                      \HR_{1,1,1,0,1}-\frac{178807}{261}
                      \HR_{1,1,1,1,0}\\
  \bar{D}_M^{(2)}  = & 32 \HR_{1,0,-1,-1,1,0}+32 \HR_{1,0,-1,1,-1,0}-8 \HR_{1,0,-1,1,1,1}+32 \HR_{1,0,1,-1,-1,1}\nonumber\\
                    & +32 \HR_{1,0,1,-1,1,-1}+\frac{64}{3} \HR_{1,0,1,-1,1,0}-72
                      \HR_{1,0,1,-1,1,1}-\frac{148}{9} \HR_{1,0,1,0,1,1}\nonumber\\
                    & +64 \HR_{1,0,1,1,-1,-1}+\frac{224}{3} \HR_{1,0,1,1,-1,0}-168 \HR_{1,0,1,1,-1,1}-\frac{1529}{9}
                      \HR_{1,0,1,1,0,1}\nonumber\\
                    & -296 \HR_{1,0,1,1,1,-1}-479 \HR_{1,0,1,1,1,0}+\frac{160}{3} \HR_{1,1,0,-1,1,1}+64 \HR_{1,1,0,1,-1,-1}\nonumber\\
                    & -\frac{8869088
                      \HR_{1,1,0,1,-1,0}}{136107}+\frac{1676416 \HR_{1,1,0,1,-1,1}}{45369}-\frac{1970464 \HR_{1,1,0,1,1,-1}}{45369}\nonumber\\
                    & +\frac{1868992 \HR_{1,1,1,-1,-1,0}}{45369}-\frac{1868992
                      \HR_{1,1,1,-1,-1,1}}{45369}-\frac{2255680 \HR_{1,1,1,-1,0,0}}{136107}\nonumber\\
                    & +\frac{451136 \HR_{1,1,1,-1,0,1}}{15123}-\frac{1868992 \HR_{1,1,1,-1,1,-1}}{45369}+\frac{322240
                      \HR_{1,1,1,-1,1,0}}{136107}\nonumber\\
                    & -\frac{708928 \HR_{1,1,1,-1,1,1}}{45369}-\frac{2126176 \HR_{1,1,1,0,-1,0}}{5041}+\frac{27111848 \HR_{1,1,1,0,-1,1}}{45369}\nonumber\\
                    & +\frac{262986220
                      \HR_{1,1,1,0,0,1}}{1224963}+\frac{33592904 \HR_{1,1,1,0,1,-1}}{45369}+\frac{1398989602 \HR_{1,1,1,0,1,0}}{1224963}\nonumber\\
                    & -\frac{7475968 \HR_{1,1,1,1,-1,-1}}{45369}-\frac{2771264
                      \HR_{1,1,1,1,-1,0}}{45369}-\frac{837824 \HR_{1,1,1,1,-1,1}}{45369}\nonumber\\
                    & +\frac{92477504 \HR_{1,1,1,1,0,-1}}{45369}+\frac{49177327 \HR_{1,1,1,1,0,0}}{20164}+\frac{322240
                      \HR_{1,1,1,1,1,-1}}{5041}\nonumber\\
                    & +\frac{5318919501463 \HR_{1,1,1,1,1,1}}{34298964}
\end{align}

\begin{align}
  \bar{D}_N^{(0)} = & -\frac{72}{11} \HR_{1,-1,1,0}+\frac{180}{11} \HR_{1,-1,1,1}+\frac{60}{11} \HR_{1,0,1,0}\nonumber\\
                    & -\frac{144}{11} \HR_{1,1,-1,0}+\frac{360}{11} \HR_{1,1,-1,1}+\frac{540}{11}
                      \HR_{1,1,1,-1}-\frac{10839}{55} \HR_{1,1,1,1}\\
  \bar{D}_N^{(1)} = & -\frac{328}{29} \HR_{1,1,-1,0,0}+\frac{428}{29} \HR_{1,1,-1,0,1}+\frac{428}{29} \HR_{1,1,-1,1,0}-\frac{528}{29} \HR_{1,1,-1,1,1}\nonumber\\
                    & -\frac{451079
                      \HR_{1,1,0,1,1}}{5046}+\frac{1684}{29} \HR_{1,1,1,-1,0}-\frac{2184}{29} \HR_{1,1,1,-1,1}-\frac{4056811 \HR_{1,1,1,0,1}}{15138}\nonumber\\
                    & -192 \HR_{1,1,1,1,-1}-\frac{3533174
                      \HR_{1,1,1,1,0}}{7569}\\
  \bar{D}_N^{(2)} = & -\frac{232020872 \HR_{1,1,0,1,-1,0}}{35287}+\frac{242178240 \HR_{1,1,0,1,-1,1}}{35287}+\frac{484356480 \HR_{1,1,0,1,1,-1}}{35287}\nonumber\\
                    & +\frac{110839452
                      \HR_{1,1,1,-1,-1,0}}{35287}-\frac{110839452 \HR_{1,1,1,-1,-1,1}}{35287}-\frac{6313068 \HR_{1,1,1,-1,0,0}}{5041}\nonumber\\
                    & +\frac{11391752 \HR_{1,1,1,-1,0,1}}{5041}-\frac{110839452
                      \HR_{1,1,1,-1,1,-1}}{35287}+\frac{5849296 \HR_{1,1,1,-1,1,0}}{35287}\nonumber\\
                    & -\frac{41400084 \HR_{1,1,1,-1,1,1}}{35287}-\frac{691680096 \HR_{1,1,1,0,-1,0}}{35287}+\frac{103164600
                      \HR_{1,1,1,0,-1,1}}{5041}\nonumber\\
                    & -\frac{321226763 \HR_{1,1,1,0,0,1}}{317583}+\frac{1448686920 \HR_{1,1,1,0,1,-1}}{35287}-\frac{347521883 \HR_{1,1,1,0,1,0}}{317583}\nonumber\\
                    & -\frac{443357808
                      \HR_{1,1,1,1,-1,-1}}{35287}-\frac{166918020 \HR_{1,1,1,1,-1,0}}{35287}-\frac{46386708 \HR_{1,1,1,1,-1,1}}{35287}\nonumber\\
                    & +\frac{412658400 \HR_{1,1,1,1,0,-1}}{5041}-\frac{3381227415
                      \HR_{1,1,1,1,0,0}}{282296}+\frac{182067300
                      \HR_{1,1,1,1,1,-1}}{35287}\nonumber\\
                    & +\frac{21952587521807
                      \HR_{1,1,1,1,1,1}}{17784648}
\end{align}

\begin{align}
  \bar{E}_3^{(-1)} & = 6 \HI_{1,0}\\
  \bar{E}_3^{(0)} & = -12 \HI_{-1,1,0}-12 \HI_{1,-1,0}+3 \HI_{1,1,1}\\
  \bar{E}_3^{(1)} & = 24 \HI_{-1,-1,1,0}+24 \HI_{-1,1,-1,0}-6
                    \HI_{-1,1,1,1}+24 \HI_{1,-1,-1,1}+24
                    \HI_{1,-1,1,-1}\nonumber\\
                   & +16 \HI_{1,-1,1,0}-54 \HI_{1,-1,1,1}-\frac{37}{3} \HI_{1,0,1,1}+48
                     \HI_{1,1,-1,-1}+56 \HI_{1,1,-1,0}\nonumber\\
                   &-126 \HI_{1,1,-1,1}-\frac{1529}{12}
                     \HI_{1,1,0,1}-222 \HI_{1,1,1,-1} -\frac{1437}{4} \HI_{1,1,1,0}\\
  \bar{E}_3^{(2)} & = -48 \HI_{-1,-1,-1,1,0}-48 \HI_{-1,-1,1,-1,0}+12
                    \HI_{-1,-1,1,1,1}-48 \HI_{-1,1,-1,-1,1}\nonumber\\
                   & -48 \HI_{-1,1,-1,1,-1} -32 \HI_{-1,1,-1,1,0}+108
                     \HI_{-1,1,-1,1,1}+\frac{74}{3} \HI_{-1,1,0,1,1}\nonumber\\
                   & -96 \HI_{-1,1,1,-1,-1}-112 \HI_{-1,1,1,-1,0}+252 \HI_{-1,1,1,-1,1}+\frac{1529}{6} \HI_{-1,1,1,0,1}\nonumber\\
                   & +444
                     \HI_{-1,1,1,1,-1}+\frac{1437}{2} \HI_{-1,1,1,1,0}-48 \HI_{1,-1,-1,-1,1}-48 \HI_{1,-1,-1,1,-1}\nonumber\\
                   & -32 \HI_{1,-1,-1,1,0}-112 \HI_{1,-1,1,-1,0}+96 \HI_{1,-1,1,-1,1}-80
                     \HI_{1,-1,1,0,-1}\nonumber\\
                   & +144 \HI_{1,-1,1,1,-1}+\frac{11841}{4} \HI_{1,-1,1,1,1}+90 \HI_{1,0,-1,1,1}-80 \HI_{1,0,1,-1,-1}\nonumber\\
                   & +130 \HI_{1,0,1,-1,1}+\frac{8105}{36}
                     \HI_{1,0,1,0,1}+250 \HI_{1,0,1,1,-1}+\frac{21503}{36} \HI_{1,0,1,1,0}\nonumber\\
                   & +96 \HI_{1,1,-1,-1,-1}-172 \HI_{1,1,-1,-1,1}+\frac{1725367 \HI_{1,1,-1,0,0}}{4161}-\frac{412061
                     \HI_{1,1,-1,0,1}}{8322}\nonumber\\
                   & -268 \HI_{1,1,-1,1,-1}-\frac{1334797 \HI_{1,1,-1,1,0}}{2774}+\frac{14659325 \HI_{1,1,-1,1,1}}{2774}+\frac{771}{2} \HI_{1,1,0,-1,1}\nonumber\\
                   & -\frac{5253406
                     \HI_{1,1,0,0,-1}}{12483}+\frac{1081050 \HI_{1,1,0,1,-1}}{1387}-1112 \HI_{1,1,1,-1,-1}-\frac{1316324 \HI_{1,1,1,-1,0}}{1387}\nonumber\\
                   & +\frac{40815981
                     \HI_{1,1,1,-1,1}}{5548}+\frac{1357934}{657} \HI_{1,1,1,0,-1}-\frac{1768682807 \HI_{1,1,1,0,0}}{898776}\nonumber\\
                   & +\frac{14459841 \HI_{1,1,1,1,-1}}{1387}-\frac{3591211175
                     \HI_{1,1,1,1,1}}{112347}\\
  \bar{E}_3^{(3)} & = 96 \HI_{-1,-1,-1,-1,1,0}+96 \HI_{-1,-1,-1,1,-1,0}-24 \HI_{-1,-1,-1,1,1,1}+96 \HI_{-1,-1,1,-1,-1,1}\nonumber\\
                   & +96 \HI_{-1,-1,1,-1,1,-1}+64 \HI_{-1,-1,1,-1,1,0}-216
                     \HI_{-1,-1,1,-1,1,1}-\frac{148}{3} \HI_{-1,-1,1,0,1,1}\nonumber\\
                   & +192 \HI_{-1,-1,1,1,-1,-1}+224 \HI_{-1,-1,1,1,-1,0}-504 \HI_{-1,-1,1,1,-1,1}-\frac{1529}{3}
                     \HI_{-1,-1,1,1,0,1}\nonumber\\
                   & -888 \HI_{-1,-1,1,1,1,-1}-1437 \HI_{-1,-1,1,1,1,0}+96 \HI_{-1,1,-1,-1,-1,1}+96 \HI_{-1,1,-1,-1,1,-1}\nonumber\\
                   & +64 \HI_{-1,1,-1,-1,1,0}+224
                     \HI_{-1,1,-1,1,-1,0}-192 \HI_{-1,1,-1,1,-1,1}+160 \HI_{-1,1,-1,1,0,-1}\nonumber\\
                   & -288 \HI_{-1,1,-1,1,1,-1}-\frac{11841}{2} \HI_{-1,1,-1,1,1,1}-180 \HI_{-1,1,0,-1,1,1}+160
                     \HI_{-1,1,0,1,-1,-1}\nonumber\\
                   & -260 \HI_{-1,1,0,1,-1,1}-\frac{8105}{18} \HI_{-1,1,0,1,0,1}-500 \HI_{-1,1,0,1,1,-1}-\frac{21503}{18} \HI_{-1,1,0,1,1,0}\nonumber\\
                   & -192
                     \HI_{-1,1,1,-1,-1,-1}+344 \HI_{-1,1,1,-1,-1,1}-\frac{3450734 \HI_{-1,1,1,-1,0,0}}{4161}+\frac{412061 \HI_{-1,1,1,-1,0,1}}{4161}\nonumber\\
                   & +536 \HI_{-1,1,1,-1,1,-1}+\frac{1334797
                     \HI_{-1,1,1,-1,1,0}}{1387}-\frac{14659325 \HI_{-1,1,1,-1,1,1}}{1387}-771 \HI_{-1,1,1,0,-1,1}\nonumber\\
                   & +\frac{10506812 \HI_{-1,1,1,0,0,-1}}{12483}-\frac{2162100
                     \HI_{-1,1,1,0,1,-1}}{1387}+2224 \HI_{-1,1,1,1,-1,-1}\nonumber\\
                   & +\frac{2632648 \HI_{-1,1,1,1,-1,0}}{1387}-\frac{40815981 \HI_{-1,1,1,1,-1,1}}{2774}-\frac{2715868}{657}
                     \HI_{-1,1,1,1,0,-1}\nonumber\\
                   & +\frac{1768682807 \HI_{-1,1,1,1,0,0}}{449388}-\frac{28919682 \HI_{-1,1,1,1,1,-1}}{1387}+\frac{7182422350 \HI_{-1,1,1,1,1,1}}{112347}\nonumber\\
                   & +96
                     \HI_{1,-1,-1,-1,-1,1}+96 \HI_{1,-1,-1,-1,1,-1}+64 \HI_{1,-1,-1,-1,1,0}+224 \HI_{1,-1,-1,1,-1,0}\nonumber\\
                   & -192 \HI_{1,-1,-1,1,-1,1}+160 \HI_{1,-1,-1,1,0,-1}-288
                     \HI_{1,-1,-1,1,1,-1}-\frac{11841}{2} \HI_{1,-1,-1,1,1,1}\nonumber\\
                   & +248 \HI_{1,-1,1,-1,-1,1}+440 \HI_{1,-1,1,-1,1,-1}+\frac{2590}{3} \HI_{1,-1,1,-1,1,0}-10608
                     \HI_{1,-1,1,-1,1,1}\nonumber\\
                   & -320 \HI_{1,-1,1,0,-1,-1}-\frac{759843 \HI_{1,-1,1,0,1,-1}}{1387}+\frac{13919645 \HI_{1,-1,1,0,1,1}}{12483}+1520 \HI_{1,-1,1,1,-1,-1}\nonumber\\
                   & +\frac{7073822
                     \HI_{1,-1,1,1,-1,0}}{4161}-\frac{40321721 \HI_{1,-1,1,1,-1,1}}{2774}+\frac{6271329955 \HI_{1,-1,1,1,0,1}}{898776}\nonumber\\
                   & -\frac{26862395 \HI_{1,-1,1,1,1,-1}}{1387}+\frac{6038799457
                     \HI_{1,-1,1,1,1,0}}{299592}-160 \HI_{1,0,1,-1,-1,-1}\nonumber\\
                   & -\frac{277814945 \HI_{1,0,1,-1,1,0}}{137313}+\frac{212581060 \HI_{1,0,1,-1,1,1}}{45771}-\frac{585}{2}
                     \HI_{1,0,1,0,-1,1}\nonumber\\
                   & -\frac{585}{2} \HI_{1,0,1,0,1,-1}+\frac{2272287250 \HI_{1,0,1,0,1,0}}{1235817}+\frac{1280}{3} \HI_{1,0,1,1,-1,-1}\nonumber\\
                   & -\frac{221375860
                     \HI_{1,0,1,1,-1,0}}{137313}+\frac{4419865}{803} \HI_{1,0,1,1,-1,1}-\frac{2624160 \HI_{1,0,1,1,0,-1}}{1387}\nonumber\\
                   & +\frac{86319175 \HI_{1,0,1,1,1,-1}}{45771}-\frac{7230866663213
                     \HI_{1,0,1,1,1,1}}{74149020}+192 \HI_{1,1,-1,-1,-1,-1} \nonumber\\
                   &+152 \HI_{1,1,-1,-1,-1,1}+344 \HI_{1,1,-1,-1,1,-1}+\frac{2590}{3} \HI_{1,1,-1,-1,1,0} -\frac{30479}{3}
                     \HI_{1,1,-1,-1,1,1}\nonumber\\
                   &-\frac{426963 \HI_{1,1,-1,0,1,-1}}{1387}+944 \HI_{1,1,-1,1,-1,-1} +1601 \HI_{1,1,-1,1,-1,0}-\frac{83605}{6} \HI_{1,1,-1,1,-1,1}\nonumber\\
                   &+\frac{2419561}{657}
                     \HI_{1,1,-1,1,0,-1}+\frac{837579887 \HI_{1,1,-1,1,0,1}}{112347}-\frac{73453112 \HI_{1,1,-1,1,1,-1}}{4161}\nonumber\\
                   &+\frac{24602317907 \HI_{1,1,-1,1,1,0}}{898776}-\frac{2981940259
                     \HI_{1,1,-1,1,1,1}}{224694}-571 \HI_{1,1,0,-1,-1,1}\nonumber\\
                   &-\frac{1218940 \HI_{1,1,0,-1,1,-1}}{1387}-\frac{1284799676 \HI_{1,1,0,-1,1,0}}{411939}+\frac{172606885810943423
                     \HI_{1,1,0,1,-1,-1}}{31412461173192}\nonumber\\
                   &+\frac{330844112035157 \HI_{1,1,0,1,-1,0}}{387808162632}-\frac{335938955928166355 \HI_{1,1,0,1,-1,1}}{62824922346384}\nonumber\\
                   &-\frac{9666835700761579
                     \HI_{1,1,0,1,1,-1}}{296343973332}-784 \HI_{1,1,1,-1,-1,-1}+\frac{32163144510925889 \HI_{1,1,1,-1,-1,0}}{1903785525648}\nonumber\\
                   &-\frac{52473680391301577
                     \HI_{1,1,1,-1,-1,1}}{1903785525648}+\frac{910517365657338137 \HI_{1,1,1,-1,0,-1}}{62824922346384}\nonumber\\
                   &+\frac{263100146907443657
                     \HI_{1,1,1,-1,0,0}}{17134069730832}-\frac{46280134588267057 \HI_{1,1,1,-1,0,1}}{188474767039152}\nonumber\\
                   &-\frac{57799520399301857 \HI_{1,1,1,-1,1,-1}}{1903785525648}+\frac{931530505381579307
                     \HI_{1,1,1,-1,1,0}}{62824922346384}\nonumber\\
                   &+\frac{1778265423341603573 \HI_{1,1,1,-1,1,1}}{62824922346384}+\frac{161624158645205183
                     \HI_{1,1,1,0,-1,-1}}{10470820391064}\nonumber\\
                   &-\frac{67738084863003155 \HI_{1,1,1,0,-1,0}}{5235410195532}-\frac{25709636977147441 \HI_{1,1,1,0,-1,1}}{2326848975792}\nonumber\\
                   &-\frac{1201750627073139241
                     \HI_{1,1,1,0,0,-1}}{205608836769984}\nonumber\\
                   &+\frac{37123461223687789039459901 \HI_{1,1,1,0,0,0}}{489612210823627499520}-\frac{15065768752177753
                     \HI_{1,1,1,0,1,-1}}{183698603352}\nonumber\\
                   &+\frac{5890725611419818825578737 \HI_{1,1,1,0,1,1}}{25769063727559342080}-\frac{2641792499452155803
                     \HI_{1,1,1,1,-1,-1}}{31412461173192}\nonumber\\
                   &-\frac{232062550415237737 \HI_{1,1,1,1,-1,0}}{3306574860336}+\frac{709822530971820472
                     \HI_{1,1,1,1,-1,1}}{3926557646649}\nonumber\\
                   &-\frac{114034845156029701411 \HI_{1,1,1,1,0,-1}}{1130848602234912}+\frac{497108468757262704945911459
                     \HI_{1,1,1,1,0,1}}{979224421647254999040}\nonumber\\
                   &+\frac{19072578396451593155
                     \HI_{1,1,1,1,1,-1}}{31412461173192}+\frac{160662586785011672453801383
                     \HI_{1,1,1,1,1,0}}{326408140549084999680}
\end{align}

\begin{align}
  \bar{BN}^{(1)} = & 112 \HR_{1,1,0}\\
  \bar{BN}^{(2)} = & -\frac{2304}{11} \HR_{1,-1,1,0}+\frac{5760}{11} \HR_{1,-1,1,1}+\frac{1920}{11} \HR_{1,0,1,0}-\frac{4608}{11} \HR_{1,1,-1,0}\nonumber\\
                   & +\frac{11520}{11} \HR_{1,1,-1,1}+\frac{17280}{11}
                     \HR_{1,1,1,-1}-\frac{593952}{55} \HR_{1,1,1,1}\\
  \bar{BN}^{(3)} = & -\frac{36864}{29} \HR_{1,1,-1,0,0}+\frac{52992}{29} \HR_{1,1,-1,0,1}+\frac{52992}{29} \HR_{1,1,-1,1,0}-\frac{69120}{29} \HR_{1,1,-1,1,1}\nonumber\\
                   & -\frac{3282240}{841}
                     \HR_{1,1,0,1,1}+\frac{223488}{29} \HR_{1,1,1,-1,0}-\frac{304128}{29} \HR_{1,1,1,-1,1}-\frac{9820736}{841} \HR_{1,1,1,0,1}\nonumber\\
                   & -27648 \HR_{1,1,1,1,-1}-\frac{19524608}{841}
                     \HR_{1,1,1,1,0}n\\
  \bar{BN}^{(4)} = & -\frac{1303119360 \HR_{1,1,0,1,-1,0}}{5041}+\frac{1354682880 \HR_{1,1,0,1,-1,1}}{5041}+\frac{2709365760 \HR_{1,1,0,1,1,-1}}{5041}\nonumber\\
                   & +\frac{613018368
                     \HR_{1,1,1,-1,-1,0}}{5041}-\frac{613018368 \HR_{1,1,1,-1,-1,1}}{5041}-\frac{208168704 \HR_{1,1,1,-1,0,0}}{5041}\nonumber\\
                   & +\frac{388641024 \HR_{1,1,1,-1,0,1}}{5041}-\frac{613018368
                     \HR_{1,1,1,-1,1,-1}}{5041}-\frac{20037888 \HR_{1,1,1,-1,1,0}}{5041}\nonumber\\
                   & -\frac{160434432 \HR_{1,1,1,-1,1,1}}{5041}-\frac{3867098112 \HR_{1,1,1,0,-1,0}}{5041}+\frac{4021788672
                     \HR_{1,1,1,0,-1,1}}{5041}\nonumber\\
                   & -\frac{181188480 \HR_{1,1,1,0,0,1}}{5041}+\frac{8085837312 \HR_{1,1,1,0,1,-1}}{5041}-\frac{209361792 \HR_{1,1,1,0,1,0}}{5041}\nonumber\\
                   & -\frac{2452073472
                     \HR_{1,1,1,1,-1,-1}}{5041}-\frac{1197317376 \HR_{1,1,1,1,-1,0}}{5041}+\frac{114483456 \HR_{1,1,1,1,-1,1}}{5041}\nonumber\\
                   & +\frac{16087154688 \HR_{1,1,1,1,0,-1}}{5041}-\frac{2378746336
                     \HR_{1,1,1,1,0,0}}{5041}+\frac{2176761600
                     \HR_{1,1,1,1,1,-1}}{5041}\nonumber\\
                   & +\frac{249395302624
                     \HR_{1,1,1,1,1,1}}{5041}
\end{align}

\begin{align}
  \bar{BN}_1^{(0)} = & 4 \HI_{1,0}\\
  \bar{BN}_1^{(1)} = & -24 \HI_{-1,1,0}-24 \HI_{1,-1,0}+78 \HI_{1,1,1}\\
  \bar{BN}_1^{(2)} = & 144 \HI_{-1,-1,1,0}+144 \HI_{-1,1,-1,0}-468
                       \HI_{-1,1,1,1}+144 \HI_{1,-1,-1,1}+144
                       \HI_{1,-1,1,-1}\nonumber\\
                     & +96 \HI_{1,-1,1,0}-756 \HI_{1,-1,1,1}+142 \HI_{1,0,1,1}+288
                       \HI_{1,1,-1,-1}+336 \HI_{1,1,-1,0}\nonumber\\
                     & -1188 \HI_{1,1,-1,1}+\frac{199}{2}
                       \HI_{1,1,0,1}-1764 \HI_{1,1,1,-1} -\frac{999}{2} \HI_{1,1,1,0}\\
  \bar{BN}_1^{(3)} & = -864 \HI_{-1,-1,-1,1,0}-864
                     \HI_{-1,-1,1,-1,0} +2808 \HI_{-1,-1,1,1,1} -864 \HI_{-1,1,-1,-1,1}\nonumber\\
                     & -864 \HI_{-1,1,-1,1,-1}-576 \HI_{-1,1,-1,1,0}
                       +4536
                       \HI_{-1,1,-1,1,1}-852 \HI_{-1,1,0,1,1}\nonumber\\
                     & -1728 \HI_{-1,1,1,-1,-1}-2016 \HI_{-1,1,1,-1,0}+7128 \HI_{-1,1,1,-1,1}-597 \HI_{-1,1,1,0,1}\nonumber\\
                     & +10584 \HI_{-1,1,1,1,-1}+2997
                       \HI_{-1,1,1,1,0}-864 \HI_{1,-1,-1,-1,1}-864 \HI_{1,-1,-1,1,-1}\nonumber\\
                     & -576 \HI_{1,-1,-1,1,0}-2016 \HI_{1,-1,1,-1,0}+1728 \HI_{1,-1,1,-1,1}-1440 \HI_{1,-1,1,0,-1}\nonumber\\
                     & +2592
                       \HI_{1,-1,1,1,-1}+\frac{41337}{2} \HI_{1,-1,1,1,1}+3780 \HI_{1,0,-1,1,1}-1440 \HI_{1,0,1,-1,-1}\nonumber\\
                     & +4500 \HI_{1,0,1,-1,1}+\frac{1625}{2} \HI_{1,0,1,0,1}+6660
                       \HI_{1,0,1,1,-1}+\frac{5903}{2} \HI_{1,0,1,1,0}\nonumber\\
                     & +1728 \HI_{1,1,-1,-1,-1}-5688 \HI_{1,1,-1,-1,1}+\frac{3566346 \HI_{1,1,-1,0,0}}{1387}+\frac{2763273
                       \HI_{1,1,-1,0,1}}{1387}\nonumber\\
                     & -7416 \HI_{1,1,-1,1,-1}-\frac{1422693 \HI_{1,1,-1,1,0}}{1387}+\frac{54624501 \HI_{1,1,-1,1,1}}{1387}+8667 \HI_{1,1,0,-1,1}\nonumber\\
                     & -\frac{5422460
                       \HI_{1,1,0,0,-1}}{1387}+\frac{17237556 \HI_{1,1,0,1,-1}}{1387}-25200 \HI_{1,1,1,-1,-1}\nonumber\\
                     & -\frac{5703624 \HI_{1,1,1,-1,0}}{1387}+\frac{174917637
                       \HI_{1,1,1,-1,1}}{2774}+\frac{1624492}{73} \HI_{1,1,1,0,-1}\nonumber\\
                     & -\frac{400772087 \HI_{1,1,1,0,0}}{49932}+\frac{158327730 \HI_{1,1,1,1,-1}}{1387}-\frac{3279409000
                       \HI_{1,1,1,1,1}}{12483}\\
  \bar{BN}_1^{(4)} = & 5184 \HI_{-1,-1,-1,-1,1,0}+5184 \HI_{-1,-1,-1,1,-1,0}-16848 \HI_{-1,-1,-1,1,1,1}\nonumber\\
                     & +5184 \HI_{-1,-1,1,-1,-1,1}+5184 \HI_{-1,-1,1,-1,1,-1}+3456 \HI_{-1,-1,1,-1,1,0}\nonumber\\
                     & -27216
                       \HI_{-1,-1,1,-1,1,1}+5112 \HI_{-1,-1,1,0,1,1}+10368 \HI_{-1,-1,1,1,-1,-1}\nonumber\\
                     & +12096 \HI_{-1,-1,1,1,-1,0}-42768 \HI_{-1,-1,1,1,-1,1}+3582 \HI_{-1,-1,1,1,0,1}\nonumber\\
                     & -63504
                       \HI_{-1,-1,1,1,1,-1}-17982 \HI_{-1,-1,1,1,1,0}+5184 \HI_{-1,1,-1,-1,-1,1}\nonumber\\
                     & +5184 \HI_{-1,1,-1,-1,1,-1}+3456 \HI_{-1,1,-1,-1,1,0}+12096 \HI_{-1,1,-1,1,-1,0}\nonumber\\
                     & -10368
                       \HI_{-1,1,-1,1,-1,1}+8640 \HI_{-1,1,-1,1,0,-1}-15552 \HI_{-1,1,-1,1,1,-1}\nonumber\\
                     & -124011 \HI_{-1,1,-1,1,1,1}-22680 \HI_{-1,1,0,-1,1,1}+8640 \HI_{-1,1,0,1,-1,-1}\nonumber\\
                     & -27000
                       \HI_{-1,1,0,1,-1,1}-4875 \HI_{-1,1,0,1,0,1}-39960 \HI_{-1,1,0,1,1,-1}\nonumber\\
                     & -17709 \HI_{-1,1,0,1,1,0}-10368 \HI_{-1,1,1,-1,-1,-1}+34128 \HI_{-1,1,1,-1,-1,1}\nonumber\\
                     & -\frac{21398076
                       \HI_{-1,1,1,-1,0,0}}{1387}-\frac{16579638 \HI_{-1,1,1,-1,0,1}}{1387}+44496 \HI_{-1,1,1,-1,1,-1}\nonumber\\
                     & +\frac{8536158 \HI_{-1,1,1,-1,1,0}}{1387}-\frac{327747006
                       \HI_{-1,1,1,-1,1,1}}{1387}-52002 \HI_{-1,1,1,0,-1,1}\nonumber\\
                     & +\frac{32534760
                       \HI_{-1,1,1,0,0,-1}}{1387}- \frac{103425336
                       \HI_{-1,1,1,0,1,-1}}{1387}+151200 \HI_{-1,1,1,1,-1,-1}\nonumber\\
                     & +\frac{34221744 \HI_{-1,1,1,1,-1,0}}{1387}-\frac{524752911 \HI_{-1,1,1,1,-1,1}}{1387}-\frac{9746952}{73} \HI_{-1,1,1,1,0,-1}\nonumber\\
                     & +\frac{400772087
                       \HI_{-1,1,1,1,0,0}}{8322}-\frac{949966380 \HI_{-1,1,1,1,1,-1}}{1387}+\frac{6558818000 \HI_{-1,1,1,1,1,1}}{4161}\nonumber\\
                     & +5184 \HI_{1,-1,-1,-1,-1,1}+5184 \HI_{1,-1,-1,-1,1,-1}+3456
                       \HI_{1,-1,-1,-1,1,0}\nonumber\\
                     & +12096 \HI_{1,-1,-1,1,-1,0}-10368 \HI_{1,-1,-1,1,-1,1}+8640 \HI_{1,-1,-1,1,0,-1}\nonumber\\
                     & -15552 \HI_{1,-1,-1,1,1,-1}-124011 \HI_{1,-1,-1,1,1,1}+28944
                       \HI_{1,-1,1,-1,-1,1}\nonumber\\
                     & +39312 \HI_{1,-1,1,-1,1,-1}+18108
                       \HI_{1,-1,1,-1,1,0}-290304
                       \HI_{1,-1,1,-1,1,1} \nonumber\\
                     & -17280 \HI_{1,-1,1,0,-1,-1}-\frac{13323042
                       \HI_{1,-1,1,0,1,-1}}{1387}+\frac{34731966 \HI_{1,-1,1,0,1,1}}{1387}\nonumber\\
                     & +113184 \HI_{1,-1,1,1,-1,-1}+\frac{67381020 \HI_{1,-1,1,1,-1,0}}{1387}-\frac{655388307
                       \HI_{1,-1,1,1,-1,1}}{1387}\nonumber\\
                     & +\frac{1961738995 \HI_{1,-1,1,1,0,1}}{16644}-\frac{1054843506 \HI_{1,-1,1,1,1,-1}}{1387}+\frac{1636490353 \HI_{1,-1,1,1,1,0}}{5548}\nonumber\\
                     & -8640
                       \HI_{1,0,1,-1,-1,-1}-\frac{1057484550 \HI_{1,0,1,-1,1,0}}{15257}+\frac{2896138440 \HI_{1,0,1,-1,1,1}}{15257}\nonumber\\
                     & -13635 \HI_{1,0,1,0,-1,1}-13635
                       \HI_{1,0,1,0,1,-1}+\frac{2536761620 \HI_{1,0,1,0,1,0}}{45771}\nonumber\\
                     & +23040 \HI_{1,0,1,1,-1,-1}-\frac{722452920 \HI_{1,0,1,1,-1,0}}{15257}+\frac{189139590}{803}
                       \HI_{1,0,1,1,-1,1}\nonumber\\
                     & -\frac{73887840 \HI_{1,0,1,1,0,-1}}{1387}+\frac{1521319230 \HI_{1,0,1,1,1,-1}}{15257}-\frac{1739984020721 \HI_{1,0,1,1,1,1}}{1373130}\nonumber\\
                     & +10368
                       \HI_{1,1,-1,-1,-1,-1}+23760 \HI_{1,1,-1,-1,-1,1}+34128 \HI_{1,1,-1,-1,1,-1}\nonumber\\
                     & +18108 \HI_{1,1,-1,-1,1,0}-323118 \HI_{1,1,-1,-1,1,1}+\frac{4652478
                       \HI_{1,1,-1,0,1,-1}}{1387}\nonumber\\
                     & +82080 \HI_{1,1,-1,1,-1,-1}+60534 \HI_{1,1,-1,1,-1,0}-548973 \HI_{1,1,-1,1,-1,1}\nonumber\\
                     & +\frac{7212246}{73} \HI_{1,1,-1,1,0,-1}+\frac{499549318
                       \HI_{1,1,-1,1,0,1}}{4161}
                       -\frac{1149502896 \HI_{1,1,-1,1,1,-1}}{1387}\nonumber\\
                     & +\frac{6314678771 \HI_{1,1,-1,1,1,0}}{16644}
                       -\frac{3100825687 \HI_{1,1,-1,1,1,1}}{4161}-15282
                       \HI_{1,1,0,-1,-1,1}\nonumber\\
                     & -\frac{16543656 \HI_{1,1,0,-1,1,-1}}{1387}-\frac{1961940664 \HI_{1,1,0,-1,1,0}}{15257}+\frac{64895701668594431
                       \HI_{1,1,0,1,-1,-1}}{581712243948}\nonumber\\
                     & -\frac{1378601274793377 \HI_{1,1,0,1,-1,0}}{21544897924}+\frac{125263181337378061 \HI_{1,1,0,1,-1,1}}{1163424487896}\nonumber\\
                     & -\frac{1211150098669531
                       \HI_{1,1,0,1,1,-1}}{5487851358}-11232 \HI_{1,1,1,-1,-1,-1}\nonumber\\
                     & +\frac{2392665851955041 \HI_{1,1,1,-1,-1,0}}{35255287512}-\frac{20464349954168681
                       \HI_{1,1,1,-1,-1,1}}{35255287512}\nonumber\\
                     & +\frac{189472199905473497 \HI_{1,1,1,-1,0,-1}}{1163424487896}+\frac{47207116920801641 \HI_{1,1,1,-1,0,0}}{317297587608}\nonumber\\
                     & -\frac{298649341166732401
                       \HI_{1,1,1,-1,0,1}}{3490273463688}-\frac{30404966076339713 \HI_{1,1,1,-1,1,-1}}{35255287512}\nonumber\\
                     & +\frac{181778915331736427 \HI_{1,1,1,-1,1,0}}{1163424487896}-\frac{366128024895147883
                       \HI_{1,1,1,-1,1,1}}{1163424487896}\nonumber\\
                     & +\frac{53912974502856191 \HI_{1,1,1,0,-1,-1}}{193904081316}-\frac{10304731594730675 \HI_{1,1,1,0,-1,0}}{96952040658}\nonumber\\
                     & -\frac{16995388457766673
                       \HI_{1,1,1,0,-1,1}}{43089795848}+\frac{186908888125819415 \HI_{1,1,1,0,0,-1}}{3807571051296}\nonumber\\
                     & +\frac{10276798443600238610374397
                       \HI_{1,1,1,0,0,0}}{9066892793030138880}-\frac{3306662120409625 \HI_{1,1,1,0,1,-1}}{3401825988}\nonumber\\
                     & +\frac{1299407259007733189665969
                       \HI_{1,1,1,0,1,1}}{477204883843691520}-\frac{384757595682873755 \HI_{1,1,1,1,-1,-1}}{581712243948}\nonumber\\
                     & -\frac{30219787275121801 \HI_{1,1,1,1,-1,0}}{61232867784}+\frac{228583496909607344
                       \HI_{1,1,1,1,-1,1}}{145428060987}\nonumber\\
                     & -\frac{44409253854794826595 \HI_{1,1,1,1,0,-1}}{20941640782128}+\frac{111015775290234750628966883
                       \HI_{1,1,1,1,0,1}}{18133785586060277760}\nonumber\\
                     & +\frac{2811281629788118067 \HI_{1,1,1,1,1,-1}}{581712243948}+\frac{35725379481317601118816231 \HI_{1,1,1,1,1,0}}{6044595195353425920}
\end{align}

\begin{align}
  \bar{T}_1^{(0)} & = 6 \HI_{1,0}\\
  \bar{T}_1^{(1)} & = -12 \HI_{-1,1,0}-12 \HI_{1,-1,0}+39
                    \HI_{1,1,1}\\
  \bar{T}_1^{(2)} & = 24 \HI_{-1,-1,1,0}+24 \HI_{-1,1,-1,0}-78 \HI_{-1,1,1,1}+24 \HI_{1,-1,-1,1}+24 \HI_{1,-1,1,-1}\nonumber\\
                  & +16 \HI_{1,-1,1,0}-126 \HI_{1,-1,1,1}+\frac{80}{3}
                    \HI_{1,0,1,1}+48 \HI_{1,1,-1,-1}+56 \HI_{1,1,-1,0}\nonumber\\
                  & -198 \HI_{1,1,-1,1}+\frac{127}{12} \HI_{1,1,0,1}-294 \HI_{1,1,1,-1}-\frac{409}{4} \HI_{1,1,1,0}\\
  \bar{T}_1^{(3)} & = -48 \HI_{-1,-1,-1,1,0}-48
                    \HI_{-1,-1,1,-1,0}+156 \HI_{-1,-1,1,1,1} -48
                    \HI_{-1,1,-1,-1,1}\nonumber\\
                  &  -48 \HI_{-1,1,-1,1,-1}-32 \HI_{-1,1,-1,1,0}+252
                    \HI_{-1,1,-1,1,1} -\frac{160}{3} \HI_{-1,1,0,1,1}\nonumber\\
                  & -96 \HI_{-1,1,1,-1,-1}-112
                    \HI_{-1,1,1,-1,0}+396 \HI_{-1,1,1,-1,1} -\frac{127}{6} \HI_{-1,1,1,0,1}\nonumber\\
                  & +588
                    \HI_{-1,1,1,1,-1}+\frac{409}{2} \HI_{-1,1,1,1,0}-48
                    \HI_{1,-1,-1,-1,1} -48 \HI_{1,-1,-1,1,-1}\nonumber\\
                  & -32 \HI_{1,-1,-1,1,0}-112 \HI_{1,-1,1,-1,0}+96 \HI_{1,-1,1,-1,1}-80
                    \HI_{1,-1,1,0,-1}\nonumber\\
                  & +144 \HI_{1,-1,1,1,-1}+\frac{5189}{4} \HI_{1,-1,1,1,1}+210 \HI_{1,0,-1,1,1}-80 \HI_{1,0,1,-1,-1}\nonumber\\
                  & +250 \HI_{1,0,1,-1,1}+\frac{2165}{36}
                    \HI_{1,0,1,0,1}+370 \HI_{1,0,1,1,-1}+\frac{7235}{36} \HI_{1,0,1,1,0}\nonumber\\
                  & +96 \HI_{1,1,-1,-1,-1}-316 \HI_{1,1,-1,-1,1}+\frac{634711 \HI_{1,1,-1,0,0}}{4161}+\frac{1001731
                    \HI_{1,1,-1,0,1}}{8322}\nonumber\\
                  & -412 \HI_{1,1,-1,1,-1}-\frac{203321 \HI_{1,1,-1,1,0}}{2774}+\frac{6715537 \HI_{1,1,-1,1,1}}{2774}+\frac{1011}{2} \HI_{1,1,0,-1,1}\nonumber\\
                  & -\frac{2899054
                    \HI_{1,1,0,0,-1}}{12483}+\frac{991354 \HI_{1,1,0,1,-1}}{1387}-1400 \HI_{1,1,1,-1,-1}-\frac{332028 \HI_{1,1,1,-1,0}}{1387}\nonumber\\
                  & +\frac{20832225
                    \HI_{1,1,1,-1,1}}{5548}+\frac{870038}{657} \HI_{1,1,1,0,-1}-\frac{502923119 \HI_{1,1,1,0,0}}{898776}+\frac{9081125 \HI_{1,1,1,1,-1}}{1387}\nonumber\\
                  & -\frac{7166073467
                    \HI_{1,1,1,1,1}}{449388}\\
  \bar{T}_1^{(4)} & = 96 \HI_{-1,-1,-1,-1,1,0}+96 \HI_{-1,-1,-1,1,-1,0}-312 \HI_{-1,-1,-1,1,1,1}+96 \HI_{-1,-1,1,-1,-1,1}\nonumber\\
                  & +96 \HI_{-1,-1,1,-1,1,-1}+64 \HI_{-1,-1,1,-1,1,0}-504
                    \HI_{-1,-1,1,-1,1,1}+\frac{320}{3} \HI_{-1,-1,1,0,1,1}\nonumber\\
                  & +192 \HI_{-1,-1,1,1,-1,-1}+224 \HI_{-1,-1,1,1,-1,0}-792 \HI_{-1,-1,1,1,-1,1}+\frac{127}{3}
                    \HI_{-1,-1,1,1,0,1}\nonumber\\
                  & -1176 \HI_{-1,-1,1,1,1,-1}-409 \HI_{-1,-1,1,1,1,0}+96 \HI_{-1,1,-1,-1,-1,1}+96 \HI_{-1,1,-1,-1,1,-1}\nonumber\\
                  & +64 \HI_{-1,1,-1,-1,1,0}+224
                    \HI_{-1,1,-1,1,-1,0}-192 \HI_{-1,1,-1,1,-1,1}+160 \HI_{-1,1,-1,1,0,-1}\nonumber\\
                  & -288 \HI_{-1,1,-1,1,1,-1}-\frac{5189}{2} \HI_{-1,1,-1,1,1,1}-420 \HI_{-1,1,0,-1,1,1}+160
                    \HI_{-1,1,0,1,-1,-1}\nonumber\\
                  & -500 \HI_{-1,1,0,1,-1,1}-\frac{2165}{18} \HI_{-1,1,0,1,0,1}-740 \HI_{-1,1,0,1,1,-1}-\frac{7235}{18} \HI_{-1,1,0,1,1,0}\nonumber\\
                  & -192
                    \HI_{-1,1,1,-1,-1,-1}+632 \HI_{-1,1,1,-1,-1,1}-\frac{1269422 \HI_{-1,1,1,-1,0,0}}{4161}-\frac{1001731 \HI_{-1,1,1,-1,0,1}}{4161}\nonumber\\
                  & +824 \HI_{-1,1,1,-1,1,-1}+\frac{203321
                    \HI_{-1,1,1,-1,1,0}}{1387}-\frac{6715537 \HI_{-1,1,1,-1,1,1}}{1387}-1011 \HI_{-1,1,1,0,-1,1}\nonumber\\
                  & +\frac{5798108 \HI_{-1,1,1,0,0,-1}}{12483}-\frac{1982708
                    \HI_{-1,1,1,0,1,-1}}{1387}+2800 \HI_{-1,1,1,1,-1,-1}+\frac{664056 \HI_{-1,1,1,1,-1,0}}{1387}\nonumber\\
                  & -\frac{20832225 \HI_{-1,1,1,1,-1,1}}{2774}-\frac{1740076}{657}
                    \HI_{-1,1,1,1,0,-1}+\frac{502923119 \HI_{-1,1,1,1,0,0}}{449388}\nonumber\\
                  & -\frac{18162250 \HI_{-1,1,1,1,1,-1}}{1387} +\frac{7166073467 \HI_{-1,1,1,1,1,1}}{224694}+96
                    \HI_{1,-1,-1,-1,-1,1}\nonumber\\
                  & +96 \HI_{1,-1,-1,-1,1,-1}+64 \HI_{1,-1,-1,-1,1,0}+224 \HI_{1,-1,-1,1,-1,0}-192 \HI_{1,-1,-1,1,-1,1}\nonumber\\
                  & +160 \HI_{1,-1,-1,1,0,-1}-288
                    \HI_{1,-1,-1,1,1,-1}-\frac{5189}{2} \HI_{1,-1,-1,1,1,1}+536 \HI_{1,-1,1,-1,-1,1}\nonumber\\
                  & +728 \HI_{1,-1,1,-1,1,-1}+\frac{1162}{3} \HI_{1,-1,1,-1,1,0}-5900
                    \HI_{1,-1,1,-1,1,1}-320 \HI_{1,-1,1,0,-1,-1}\nonumber\\
                  & -\frac{247571 \HI_{1,-1,1,0,1,-1}}{1387}+\frac{6935939 \HI_{1,-1,1,0,1,1}}{12483}+2096 \HI_{1,-1,1,1,-1,-1}\nonumber\\
                  & +\frac{3995630
                    \HI_{1,-1,1,1,-1,0}}{4161}-\frac{25993133 \HI_{1,-1,1,1,-1,1}}{2774}+\frac{2339932507 \HI_{1,-1,1,1,0,1}}{898776}\nonumber\\
                  & -\frac{20346339 \HI_{1,-1,1,1,1,-1}}{1387}+\frac{1987391897
                    \HI_{1,-1,1,1,1,0}}{299592}-160 \HI_{1,0,1,-1,-1,-1}\nonumber\\
                  & -\frac{190762625 \HI_{1,0,1,-1,1,0}}{137313}+\frac{172992580 \HI_{1,0,1,-1,1,1}}{45771}-\frac{585}{2}
                    \HI_{1,0,1,0,-1,1}-\frac{585}{2} \HI_{1,0,1,0,1,-1}\nonumber\\
                  & +\frac{1401897460 \HI_{1,0,1,0,1,0}}{1235817}+\frac{1280}{3} \HI_{1,0,1,1,-1,-1}-\frac{131698420
                    \HI_{1,0,1,1,-1,0}}{137313}\nonumber\\
                  & +\frac{3771385}{803} \HI_{1,0,1,1,-1,1}-\frac{13518640 \HI_{1,0,1,1,0,-1}}{12483}+\frac{94194535 \HI_{1,0,1,1,1,-1}}{45771}\nonumber\\
                  & -\frac{224239650581
                    \HI_{1,0,1,1,1,1}}{7414902}+192 \HI_{1,1,-1,-1,-1,-1}+440 \HI_{1,1,-1,-1,-1,1}\nonumber\\
                  & +632 \HI_{1,1,-1,-1,1,-1}+\frac{1162}{3} \HI_{1,1,-1,-1,1,0}-\frac{19523}{3}
                    \HI_{1,1,-1,-1,1,1}\nonumber\\
                  & +\frac{85309 \HI_{1,1,-1,0,1,-1}}{1387}+1520 \HI_{1,1,-1,1,-1,-1}+1201 \HI_{1,1,-1,1,-1,0}\nonumber\\
                  & -\frac{65065}{6} \HI_{1,1,-1,1,-1,1}+\frac{1286089}{657}
                    \HI_{1,1,-1,1,0,-1}+\frac{294745883 \HI_{1,1,-1,1,0,1}}{112347}\nonumber\\
                  & -\frac{66781712 \HI_{1,1,-1,1,1,-1}}{4161}+\frac{7672706315 \HI_{1,1,-1,1,1,0}}{898776}-\frac{1500994997
                    \HI_{1,1,-1,1,1,1}}{112347}\nonumber\\
                  & -331 \HI_{1,1,0,-1,-1,1}-\frac{373788 \HI_{1,1,0,-1,1,-1}}{1387}-\frac{1040075156 \HI_{1,1,0,-1,1,0}}{411939}\nonumber\\
                  & +\frac{71036008625022455
                    \HI_{1,1,0,1,-1,-1}}{31412461173192}-\frac{408164510625851 \HI_{1,1,0,1,-1,0}}{387808162632}\nonumber\\
                  & +\frac{112240612313169253 \HI_{1,1,0,1,-1,1}}{62824922346384}-\frac{1671079834843531
                    \HI_{1,1,0,1,1,-1}}{296343973332}\nonumber\\
                  & -208 \HI_{1,1,1,-1,-1,-1}+\frac{4630766069502545 \HI_{1,1,1,-1,-1,0}}{1903785525648}-\frac{23840913916053689
                    \HI_{1,1,1,-1,-1,1}}{1903785525648}\nonumber\\
                  & +\frac{236736287552028065 \HI_{1,1,1,-1,0,-1}}{62824922346384}+\frac{65236932450127865
                    \HI_{1,1,1,-1,0,0}}{17134069730832}\nonumber\\
                  & -\frac{324735078822098209 \HI_{1,1,1,-1,0,1}}{188474767039152}-\frac{33937640451327857
                    \HI_{1,1,1,-1,1,-1}}{1903785525648}\nonumber\\
                  & +\frac{217890103688771579 \HI_{1,1,1,-1,1,0}}{62824922346384}-\frac{98098907738411707 \HI_{1,1,1,-1,1,1}}{62824922346384}\nonumber\\
                  & +\frac{60053281459284215
                    \HI_{1,1,1,0,-1,-1}}{10470820391064}-\frac{14276309865069491 \HI_{1,1,1,0,-1,0}}{5235410195532}\nonumber\\
                  & -\frac{17608277669750297 \HI_{1,1,1,0,-1,1}}{2326848975792}+\frac{113893834864667471
                    \HI_{1,1,1,0,0,-1}}{205608836769984}\nonumber\\
                  & +\frac{2578864206755062092340321 \HI_{1,1,1,0,0,0}}{97922442164725499904}-\frac{4028157988084297
                    \HI_{1,1,1,0,1,-1}}{183698603352}\nonumber\\
                  & +\frac{342310613376007581255845 \HI_{1,1,1,0,1,1}}{5153812745511868416}-\frac{580499376653049347
                    \HI_{1,1,1,1,-1,-1}}{31412461173192}\nonumber\\
                  & -\frac{47970936019094809 \HI_{1,1,1,1,-1,0}}{3306574860336}+\frac{175367882302986940
                    \HI_{1,1,1,1,-1,1}}{3926557646649}\nonumber\\
                  & -\frac{48827026671130603771 \HI_{1,1,1,1,0,-1}}{1130848602234912}+\frac{29301329002972334771885311
                    \HI_{1,1,1,1,0,1}}{195844884329450999808}\nonumber\\
                  & +\frac{4274072449929751295 \HI_{1,1,1,1,1,-1}}{31412461173192}+\frac{9649675317039589744890323 \HI_{1,1,1,1,1,0}}{65281628109816999936}
\end{align}


\bibliographystyle{JHEP}
\bibliography{vac3}
\end{document}